\documentclass{article}

\usepackage{arxiv}
\RequirePackage[colorlinks,citecolor=blue,urlcolor=blue,allcolors=blue]{hyperref}
\usepackage[utf8]{inputenc} % allow utf-8 input
\usepackage[T1]{fontenc}    % use 8-bit T1 fonts
\usepackage{url}            % simple URL typesetting
\usepackage{booktabs}       % professional-quality tables
\usepackage{amsfonts}       % blackboard math symbols
\usepackage{nicefrac}       % compact symbols for 1/2, etc.
\usepackage{microtype}      % microtypography
\usepackage{lipsum}		% Can be removed after putting your text content
\usepackage{graphicx}
\usepackage{apacite}
\usepackage{doi}
\usepackage{amsmath,amsthm,amssymb,scrextend,mathtools}
\usepackage{xr}
\newtheorem{theorem}{Theorem}
\newtheorem{proposition}{Proposition}

\newtheorem{definition}{Definition}
\usepackage{natbib}
\usepackage{lmodern} %%helps deal with font size problems
\usepackage{dsfont}

\usepackage{blindtext}
\usepackage{multicol}

\newcommand{\jarad}[1]{{\color{white} Jarad: #1}}

\title{Bayesian stacking via proper scoring rule optimization using a Gibbs posterior}
\author{Spencer Wadsworth \\
	University of Connecticut\\
	Storrs, Connecticut\\
	\texttt{iac25002@uconn.edu} \\
	\And
	Jarad Niemi \\
	Iowa State University\\
	Ames, Iowa\\
	\texttt{niemi@iastate.edu} \\
}

\date{}

\renewcommand{\headeright}{}
\renewcommand{\shorttitle}{Stacked Gibbs Posterior}

\begin{document}
\maketitle
\begin{abstract}
In collaborative forecast projects, the combining of multiple probabilistic 
forecasts into an ensemble is standard practice, with linear pooling
being a common combination method.
The weighting scheme of a linear pool should be 
tailored to the specific research question, and weight selection is often 
performed via optimizing a proper scoring rule. This is known as optimal
linear pooling.
Besides optimal linear pooling, Bayesian predictive synthesis has 
emerged as a model probability updating scheme which is more flexible than
standard Bayesian model averaging and which provides a Bayesian 
solution to selecting model weights for a linear pool.
In many problems, equally weighted linear pool forecasts often 
outperform forecasts constructed using sophisticated weight selection methods.
% Existing methods may or may not improve ensembles for given datasets, and 
% there is room for additional methodology.
Thus regularization to an equal weighting of foreacsts may be a valuable 
addition to any weight selection method.
In this manuscript, we introduce an optimal linear pool
based on a Gibbs posterior over stacked model weights optimized
over a proper scoring rule. The Gibbs posterior extends stacking 
into a Bayesian framework by allowing for optimal weight solutions to be 
influenced by a prior distribution, and it also provides uncertainty 
quantification of weights in the form of a probability distribution. 
We compare ensemble forecast performance with model averaging methods and equal 
weighted models in simulation studies and in a real data example from the 
2023-24 US Centers for Disease Control FluSight competition. In both the 
simulation studies and the FluSight analysis, the stacked Gibbs posterior 
produces ensemble forecasts which often outperform the ensembles of other 
methods.
\end{abstract}
% 
% 
% % keywords can be removed

\section{Introduction}
\begin{figure}[hbt!]
    \centering
    \includegraphics[scale=.65]{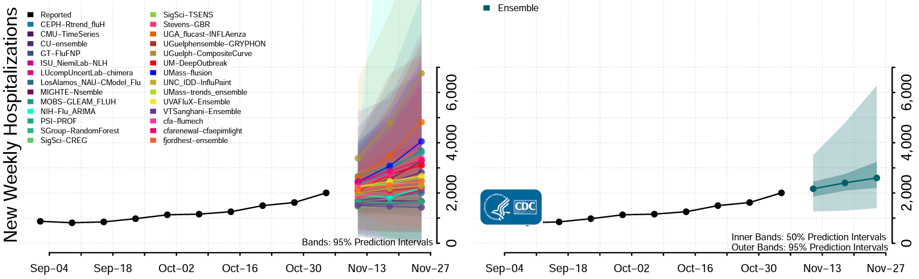}
    \caption{United States flu hospitalization forecasts from multiple 
    competing forecast teams (left) for 1-4 week ahead horizons from the week 
    of November 6, 2023 and an ensemble forecast (right) made by combining all 
    competing forecasts into one. Image downloaded from 
    https://www.cdc.gov/flu/weekly/flusight/flu-forecasts.htm}
    \label{fig:cdc_forecasts}
\end{figure}

Forecasting future events is the object of much scientific and social activity 
and informs many public and private decisions. Decision making tends to be 
better if uncertainties are attached to the forecasts 
\cite[]{ramos2013probabilistic,joslyn2012uncertainty}, and in many fields the 
focus on making forecasts probabilistic is growing 
\cite[]{wang2023forecast,hong2020energy,kapetanios2015generalised,
gneiting2014probabilistic,collins2007ensembles,palmer2002economic}. 
It's rare that nature's data generating processes can be known, so a common 
forecasting approach is to produce multiple statistical models or predictive 
procedures each targeting the same event. One may then search for and select 
the best forecasts or in some way combine forecasts to improve overall 
forecast performance
\cite[]{yao2018using,clyde2013bayesian,biggerstaff2016results,
bernardo1994bayesian}. Large forecast competitions and official forecast 
hubs exist to exploit the skill of multiple forecasts and forecasters 
\cite[]{mathis2024evaluation,Cramer2022-hub-dataset,hyndman2020brief,
makridakis2020m4,reich2019collaborative,biggerstaff2016results,
hong2016probabilistic}. In these initiatives, multiple forecasters submit 
separate forecasts targeting the same event, and the performances of the 
forecasts -- typically assessed via a \textit{proper scoring rule} -- are 
compared.
Combining multiple candidate forecasts into an ensemble forecast is common 
practice in forecast hubs, and this often leads to forecasts which are 
superior to the individual forecasts \cite[]{wang2023forecast,
li2023bayesian,gyamerah2020probabilistic,li2019combining,
reich2019collaborativeens}. For example, in the 2023-24 United States Centers 
for Disease Control and Prevention (CDC) collaborative flu forecasting 
competition --also known as FluSight-- ensemble forecasts largely outperformed 
individual forecasts in predicting flu hospitalizations. In fact the 
forecast published weekly by the CDC as the official forecast is an ensemble 
of all competing forecasts \cite[]{mathis2024evaluation}. Figure 
\ref{fig:cdc_forecasts} shows an example of FluSight hospitalization 
forecasts from many competing teams on the left where on the right is shown 
the ensemble forecast constructed by combining all competing forecasts. It was
this forecast competition which motivated the work herein.

There are many methods for constructing an ensemble forecast including but not 
limited to linear and nonlinear methods \cite[]{yao2018using,
geweke2011optimal,hall2007combining,bassetti2018bayesian,
gneiting2013combining,ranjan2010combining} which combine
probability density/mass functions (PDF), cumulative distribution functions 
(CDF), or quantile functions \cite[]{lichtendahl2013better}. The focus in this 
manuscript is on combining forecasts via \textit{linear pooling}, also known as 
model stacking \cite[]{yao2018using,geweke2011optimal,hall2007combining,
stone1961opinion}. Most linear pooling combination methods involve selecting 
a set of combination weights, a point on the simplex, which optimize the 
ensemble according to some criteria but which lack any measure of weight
uncertainty in the optimization. Weights in an ensemble may also be selected 
by maximizing a likelihood or via a Bayesian posterior distribution, but when 
so constructed ensemble forecasts may not be suited for the problem specific 
criteria under which forecasts are assessed. Bayesian model probabilities may 
be used to select weights, but this may also not be well 
suited to problem specific needs. An important case of linear pooling is 
combining via a simple average over the forecasts. This equal weighting (EQW) 
scheme frequently outperforms sophisticated combination methods leading to 
what is known as the 
``forecast combination puzzle'' \cite[]{frazier2023solving,
claeskens2016forecast,smith2009simple}. 

In this manuscript, we introduce a new method which we call the stacked 
Gibbs posterior (SGP), for combining probabilistic forecasts by optimizing a 
linear pool. The SGP provides uncertainty quantification of the weights in 
the ensemble and allows the use of a prior distribution to inform or 
influence the optimized weights. The SGP is attained by constructing a Gibbs 
posterior which is a type of posterior probability distribution constructed 
not within the standard Bayesian framework but by specifying a risk function 
which one desires to optimize, assigning a prior distribution to the 
optimization parameters, and updating parameter distributions via Bayes 
theorem \cite[]{martin2022direct,bissiri2016general}. The SGP is a Gibbs 
posterior designed specifically to optimize model weights in a forecast with 
the object being to minimize a proper scoring rule 
\cite[]{gneiting2007strictly}. 

Section \ref{sec:preliminaries} includes a brief 
review of probabilistic forecasts and linear pool combinations of forecasts. 
In section \ref{sec:sgp} the SGP is developed, and important details of the 
Gibbs posterior and proper scoring rules are covered. Section 
\ref{sec:simulation} includes three simulation studies to analyze the SGP and 
compare it with other forecast combination methods. In these studies the SGP 
outperforms model averaging and EQW methods for combining forecasts. In section
\ref{sec:flu_analysis} the SGP is used to construct ensemble forecasts for the 
FluSight project where it also outperforms model averaging and EQW approaches. 
The manuscript is concluded in section \ref{sec:4conc}.

% \section{Situation}

% Consider the case where one is given data $\mathcal{Y} = (Y_1, ..., y_N)$ and a set of $K$ statistical models $\mathcal{M} = (M_1, ..., M_K)$. The intent of each model $M_k$ is to predict a new data point $\tilde{Y}$.
% In most practical situations this is an $\mathcal{M}$-open setting where none of the $K$ models is equal to the true data generating distribution. For each model, it's unknown how $Y_i$ was used to inform the model, and it's possible that additional external data were used. The goal is to create an linear ensemble model of the models in $\mathcal{M}$

% \[
%   p^{(E)}(y) = \sum_{k = 1}^K w_k p_{M_k} (y)
% \]
% by selecting weights $w_k$ which are optimal according to some scoring rule $S(p^{(E)}, y)$. $S(\cdot, \cdot)$ is chosen to be the CRPS. The weights $w_k$ are chosen as from the Gibbs posterior

% \[
% \pi^{(\eta)}(\theta) \propto \text{exp}\{-\eta n R_n(\theta)\} \pi(\theta)
% \]
% where 

% \[
%   R_n(\theta) = \frac{1}{n} CRPS(P^{(E)}, y_n)
% \]
% Here $\theta = \textbf{w} = (w_1, ..., w_k)$. 

\section{Probabilistic forecasts and linear pooling} \label{sec:preliminaries}

We consider the situation where a set of time indexed data $y_1,..., y_T$ 
is given, and one 
is tasked with producing a probabilistic forecast for the unobserved 
$\tilde{y}_{T+h}$ for $h$ time steps ahead of the latest observed time $T$. The 
forecaster produces a forecast $P$ which is probabilistic and which 
may also be indexed by $t \in \{1, ..., T\}$. $P$ may be a PDF, CDF, quantile 
function, set of quantiles or intervals, or any other representation used in 
probabilistic forecasting 
\cite[]{wadsworth2023mixture,gneiting2014probabilistic}. For the 
remainder of this paper, unless otherwise stated, we assume $P$ is the CDF 
forecast of a continuous random variable and $p$ is the corresponding PDF. 
\cite{gneiting2007probabilistic} argue that a probabilistic forecast should 
be evaluated based on sharpness subject to calibration. Sharpness refers to 
how concentrated the forecast uncertainties are, and calibration refers to 
the uniformity of the probability integral transform (PIT). 
Sharpness can be evaluated by the width of predictive intervals but is often 
measured by using proper scoring rules.
The PIT is made by evaluating the probabilistic forecast at the true value 
after it is observed. For the CDF $P$, the PIT of the observed 
$y_{T+h}$, $P(y_{T+h})$ should be roughly uniformly distributed if the 
forecast is well calibrated. The PIT is assessed by making and visualizing 
its histogram \cite[]{gneiting2007probabilistic,hamill2001interpretation}.

When in the situation where there are multiple models used to create 
probabilistic forecasts, we define the set of models as 
$\mathcal{M} = (M_1, ..., M_C)$ where $M_c$ is one model and 
$c \in (1,2,...,C)$. We say $M_{\mathcal{T}}$ is the true model which generates 
$\{y_t\}$. The setting where the true model 
$M_{\mathcal{T}} \notin \mathcal{M}$ is referred to as the $\mathcal{M}-$open 
setting \cite[]{yao2018using,bernardo1994bayesian}. In real world problems, it 
is often the case that forecasts belong to the $\mathcal{M}-$open setting. 
Combining the candidate forecasts in this setting is a 
powerful way to improve forecasts and has been done now for over half a 
century \cite[]{wang2023forecast}. 

\subsection{Linear pooling}
The most common method for combining several probabilistic forecasts is known 
as \textit{linear opinion pooling} which is a linear combination of multiple 
probabilistic forecasts. For $C$ candidate forecasts of the event 
$\tilde{y}_{t+h}$, $h$ time steps into the %future from an observed time $t$ 
with CDFs $P_{c}(\tilde{y}_{t + h})$ for $c \in (1,...,C)$, a linear pool 
is defined as the mixture distribution in 
(\ref{eq:linp_forc}). Here, the $w_c$ are linear pool weights. 
To be a proper mixture distribution, the weights must 
be constrained such that they are each nonnegative and together sum to 1, or 
$w_c \geq 0$ and $\sum_{c = 1}^C w_c = 1$.

\begin{equation}
 \label{eq:linp_forc}
 \bar{P}_{\omega}(\tilde{y}_{t + h}) = \sum_{c = 1}^C w_c P_c(\tilde{y}_{t + h})
\end{equation}
Given several individual forecasts, the goal of a forecaster in combining them 
via linear pooling is to select the vector of weights 
$\omega = (w_1, ..., w_C)$ in such a way that the ensemble forecast is 
optimized \cite[]{wang2023forecast,stone1961opinion}. 

If the weights in (\ref{eq:linp_forc}) are optimized according to some score 
function, it is known as an 
\textit{optimal prediction pool} \cite[]{geweke2011optimal,hall2007combining}. 
\cite{yao2018using} introduced a semi-Bayesian approach where a leave-one-out 
(LOO) estimator for combination weights consists of LOO posterior predictive 
distributions of candidate forecast models optimized over a proper scoring 
rule \cite[]{gneiting2007strictly}. It is common to select weights which 
minimize the logarithmic score (LogS) \cite[]{li2023bayesian,yao2018using,
geweke2011optimal,hall2007combining} or weights that minimize the continuous 
ranked probability score (CRPS) \cite[]{berrisch2023crps,li2019combining,
thorey2017online}. Both of these scores are negatively oriented proper 
scoring rules. 
More discussion on proper scoring rules is given in section \ref{sec:sgp}. 
Weight selection methods are often tailored to be problem specific, for 
example they may be formulated to be dynamic \cite[]{li2023bayesian,
billio2013time} or to vary at different points on the PDF 
\cite[]{berrisch2023crps}.

Alternatively, weights in (\ref{eq:linp_forc}) might be selected through model 
averaging which \cite{wang2023forecast} emphasize is distinct from forecast 
combination. Forecast combination is combining forecasts with the aim of 
creating an optimal forecast, whereas model averaging determines model 
probabilities and provides a measure of model uncertainty. The most common 
model averaging approach is called Bayesian model averaging (BMA). In BMA a 
prior probability, $p(\mathcal{M}_c)$, is assigned to each model, and the 
posterior model probability is updated as data is observed according to Bayes 
theorem as

\begin{equation}
    \label{eq:bma}
    p_{BMA}(\mathcal{M}_c |\boldsymbol{y}_t) = 
    \frac{p(\boldsymbol{y}_t |\mathcal{M}_c)p(\mathcal{M}_c)}{\sum_{j = 1}^C 
    p(\boldsymbol{y}_t |\mathcal{M}_j)p(\mathcal{M}_j)}
\end{equation}
where $\boldsymbol{y}_t$ represents all data observed up to time $t$. 
To forecast a future event $\tilde{y}_{t+h}$, one may formulate a predictive 
mixture distribution of 
the form of (\ref{eq:linp_forc}) where the $c^{th}$ forecast model is 
$P_c(y_{t + h}) = P(y_t | \mathcal{M}_c)$ and the weight given to the model 
is $w_c = p_{BMA}(\mathcal{M}_c | \boldsymbol{y}_t)$ 
\cite[]{raftery1996hypothesis}. In the model combination setting, a major 
drawback of BMA is that when the data is large enough, BMA assigns probability 
1 to a single candidate model and probability 0 to all other models. This is 
demonstrated in \cite{yao2018using} and in a simulation study in section 
\ref{sec:simulation} herein. 
Adaptive variable selection (AVS) is an alternative model averaging method 
introduced by \cite{lavine2021adaptive}, and it falls within the Bayesian 
predictive synthesis framework \cite[]{tallman2024bayesian,mcalinn2019dynamic}. 
In AVS, a prior distribution is assigned to each candidate model, and a Gibbs 
posterior model probability is computed for each model. In AVS, 
posterior updating is based on the exponential of some problem specific 
function, $S(\cdot)$,
% $S:\Theta \times \mathcal{Y} \rightarrow \mathbb{R}$ 
rather than on a model likelihood as in BMA. The AVS model probability is then

\begin{equation}
    \label{eq:avs}
    p_{AVS}(\mathcal{M}_c | \boldsymbol{y}_t) 
    \propto \text{exp}\{ -\eta S(\mathcal{M}_c)\} p(\mathcal{M}_c)
\end{equation}
where $\eta$ is a model tuning parameter to be selected based on the needs of 
the problem. AVS allows for much more flexibility in updating posterior model 
averages, but as we show in a simulation study 
in section \ref{sec:simulation} AVS 
may, like BMA, tend to prefer one candidate model over the others.

Among all weight selection methods for linear pooling, one of the most 
consistently well performing combinations is to give equal weight to all 
models in (\ref{eq:linp_forc}) so that $w_1 = w_2 = ... = w_C = 1/C$. This 
phenomenon has received much attention (see for example 
\cite{frazier2023solving,claeskens2016forecast,smith2009simple}), 
and fitting a simple average is obviously easier than using most weight 
optimization methods. Thus for another weight selection method to be worth 
implementing, it should outperform the EQW scheme. It has thus been recommended 
that an EQW average be used as a baseline by which other methods 
are compared 
\cite[]{li2023bayesian,clemen1989combining}.

% \begin{itemize}
%     \item Use equal weights as baseline (Li et al. 2022, Stock \& Watson 2004, Clemen 1989)
%     \item More literature on puzzle in point forecasts (Xiaoqian 2023, Frazier et al. 2024)
%     \item Minimize CRPS (Thorey et al. 2017, Berrisch & Ziel 2021, li 2019, Raftery et al. 2005)
%     \item Time dependent weights (Li et al. 2022)
%     \item Weights specific to problem (Mcalinn West 2019, Lavine 2021, Tallman & West 2024)
%     \item Bayesian (BMA, Lavine 2021, Tallman & West 2024, Yao 2018?not really Bayesian)
% \end{itemize}

\section{Stacked Gibbs posterior} \label{sec:sgp}

For weight selection of a linear pooled ensemble forecast, the stacked Gibbs 
posterior (SGP), introduced in this section, is used to estimate forecast 
combination weights $\boldsymbol{\omega}$, 
the vector of weights in the linear pool of 
(\ref{eq:linp_forc}). The SGP is defined in (\ref{eq:sgp}) and consists of 
three  components, including a prior distribution on $\boldsymbol{\omega}$, 
$\pi(\boldsymbol{\omega})$, 
an empirical risk function $S_n(\cdot)$ based on a proper scoring rule, and a 
tuning parameter $\eta$. Each of these is discussed in further detail in 
sections \ref{sec:gibb_post} and \ref{sec:prop_score}.

\begin{equation}
    \label{eq:sgp}
    \pi^{(\eta)}_n (\boldsymbol{\omega}) \propto 
    \text{exp}\{-\eta n S_n(\boldsymbol{\omega})\} 
    \pi(\boldsymbol{\omega})
\end{equation}

The SGP is a probability distribution over the simplex
which allows for inferential 
analysis of the weights. This is in contrast to the stacking done by 
\cite{yao2018using} where the weights are optimized as a point on the simplex. 
The SGP also differs from the posterior model averaging in (\ref{eq:bma}) and 
(\ref{eq:avs}) where the posterior 
probability is point assigned directly to candidate models rather than 
to the model weights. 
In the remainder of this section a general Gibbs posterior is defined in 
more detail following \cite{martin2022direct}, proper scoring rules are 
defined following \cite{gneiting2007strictly} with specific emphasis on the 
CRPS, and a posterior consistency result is presented. 

\subsection{Gibbs posterior} \label{sec:gibb_post}

The standard Bayesian posterior distribution for an unknown parameter 
$\theta$ given some data $y$ is defined by Bayes theorem as 
$\pi(\theta | y ) = c L(y|\theta) \pi(\theta)$
so that $\pi(\theta|y)$ is equal to the joint distribution of a statistical 
model $L(y|\theta)$ and a prior distribution $\pi(\theta)$ times a normalizing 
constant $c$. When the model $L(y|\theta)$ does not exist or the goal is 
inference on parameters which minimize some loss function rather than 
maximize a likelihood, the Gibbs posterior is a alternative to the standard 
posterior distribution which allows for inference on the minimizer of a 
selected function. The Gibbs posterior has been used in many applications, 
and has shown promising results, often outperforming the Bayesian posterior 
in question specific parameter estimation, prediction, model selection, and 
model averaging \cite[]{martin2022direct,loaiza2021focused,
lavine2021adaptive,syring2017gibbs,jiang2008gibbs}.

Following \cite{martin2022direct}, the setup for constructing a Gibbs posterior 
is as follows. We are given data, 
$y_1, ..., y_n \in \mathcal{Y}$ typically assumed to be independent and 
identically distributed (i.i.d.) from some inaccessible distribution 
$\mathcal{L}$. We do not assume a likelihood model for the data, either 
because one is not available or because inference via a likelihood is not 
useful for the problem. Instead we define a loss function parameterized by 
$\theta \in \Theta$, 
$l_{\theta}(y) : \Theta \times \mathcal{Y} \rightarrow \mathbb{R}$. 
Two common loss functions are the squared-error and the absolute-error losses. 
The goal for prediction is to minimize the loss function.

A risk function $R(\theta)$ is the expectation of the loss function 
$l_{\theta}(y)$ over the distribution $\mathcal{L}$, or 
$R(\theta) = E l_{\theta}(Y)$. Primarily, interest is in the risk minimizer 
$\theta^*$ defined in (\ref{eq:risk_min}) where $\in$ indicates that 
$\theta^*$ may not be unique.

\begin{equation}
    \label{eq:risk_min}
    \theta^* \in \underset{\theta \in \Theta}{\mathrm{argmin}} R(\theta)
\end{equation}
Since $R(\theta)$ is inaccessible, an empirical risk function $R_n(\theta)$ 
based on the given data is formulated in (\ref{eq:emp_risk}). Then 
$\hat{\theta}_n$, an estimate of $\theta^*$,  is estimated by the estimator 
in (\ref{eq:emp_risk_min}).  

\begin{equation}
\label{eq:emp_risk}
    R_n(\theta) = \frac{1}{n}\sum_{i = 1}^n l_{\theta}(y_i)
\end{equation}

\begin{equation}
    \label{eq:emp_risk_min}
    \hat{\theta}_n = \underset{\theta \in \Theta}{\mathrm{arg min}} R_n(\theta)
\end{equation}
The estimator can be viewed as an $M$-estimator, the statistical properties 
of which have been extensively studied \cite[]{boos2013essential, 
van1998asymptotic}.

The purpose of the Gibbs posterior is to produce probabilistic uncertainty 
for the unknown risk minimizer. In this setting, the Gibbs 
posterior is defined in (\ref{eq:gibb_post}) where $R_n(\theta)$ 
is the empirical 
risk from (\ref{eq:emp_risk}). The Gibbs posterior is proportional to the 
product of a prior distribution assigned to $\theta$ and the exponential of 
the empirical risk times a parameter $\eta > 0$ which \cite{martin2022direct} 
refer to as the \textit{learning rate}. 

\begin{equation}
    \label{eq:gibb_post}
    \pi^{(\eta)}_n (\theta) \propto \text{exp}\{-\eta n R_n(\theta)\} 
    \pi(\theta)
\end{equation}

The learning rate $\eta$ is a tuning parameter used to control the balance 
between the prior distribution and the data. It should not be considered a 
model parameter to which one may assign a prior, but rather should be 
selected by a data driven tuning or some other problem specific selection. 
Often the learning rate is tuned so that posterior credible intervals have 
frequentist coverage probability, or it is selected via cross validation to 
optimize some criteria \cite[]{martin2022direct,syring2017gibbs,
bissiri2016general,zhang2006information}. Importantly the asymptotic 
consistency of the Gibbs posterior is not dependent on $\eta$, so selecting 
$\eta$ by most well reasoned methods should be appropriate 
\cite[]{martin2022direct,loaiza2021focused}.

For the SGP in (\ref{eq:sgp}), the more general $\theta 
\in \Theta$ is replaced with $\omega \in \{w_1, ..., w_C : 0 \leq w_c \leq 1, 
\sum_c w_c = 1\}$, and we consider empirical risk functions based on one of 
two loss 
functions. The first of these functions is in (\ref{eq:er_iid}), and is used 
for the simple case of i.i.d. data $y_1,..., y_n$ and candidate models which 
are provided once and remain constant as more data become available.

\begin{equation}
    S_n(\omega) = G_n(\omega) = \frac{1}{n}\sum_{i = 1}^n 
    G(\bar{P}_{\omega}, y_i).
    \label{eq:er_iid}
\end{equation}
Here $\bar{P}_{\omega}$ is the CDF of an ensemble forecast with $C$ 
competing forecast models from (\ref{eq:linp_forc}). Each component $P_c$ 
is fully specified, so that the only parameters to be estimated are model 
weights. The function $G(\cdot, \cdot)$ is a proper scoring rule 
(see section \ref{sec:prop_score}).
A second empirical risk, used in a time dynamic setting is given in 
(\ref{eq:er_dyn}). Here we have an observed time series $y_1, ..., y_T$, and 
we want to construct an ensemble model for forecasting the future observation 
$\tilde{y}_{T+h}$. 

\begin{equation}
   S_T(\omega) = G_T(\omega) = \frac{1}{T}\sum_{t = 1}^T \alpha^{T-t} 
   G(\bar{P}_{t, \omega}, y_t)
    \label{eq:er_dyn}
\end{equation}
In this dynamic setting, $\alpha^{T-t}$ is a discount factor 
\cite[]{koop2013large,raftery2010online} used to lessen the influence of 
previous forecasts making the performance of the most recent forecasts 
the most influential. In \cite{lavine2021adaptive}, the discount factor is 
fixed at $\alpha = 0.98$, and we use the same value herein. The linear pool 
$\bar{P}_{t, \omega}$ is made up of competing models which may or may not be 
time dependent. In the second simulation in section \ref{sec:simulation}, 
the risk in (\ref{eq:er_dyn}) is used where competing models are fixed in 
time whereas in both the third simulation study and 
the analysis of flu forecasts in section 
\ref{sec:flu_analysis}, the forecasts vary in time.

\subsection{Proper scoring rules and the continuous ranked probability score} 
\label{sec:prop_score}

In this subsection, we review the definition of a proper scoring rule and 
give the definitions for the LogS and the CRPS. Also included are additional 
properties of the CRPS. 
Following \cite{gneiting2007strictly}, a scoring rule is a function 
$G:\mathcal{P} \times \mathcal{Y} \rightarrow [-\infty, \infty]$ where 
$\mathcal{P}$ is a convex class of probability measures on 
$(\mathcal{Y}, \mathcal{A})$, $\mathcal{A}$ being a $\sigma$-algebra on the 
set $\mathcal{Y}$.  $G(P, y)$ is then a score on how well a predictive 
distribution, typically a probabilistic forecast, $P \in \mathcal{P}$ 
predicts a realized value $y$ on the sample space. A negatively oriented 
scoring rule is a proper scoring rule if for two forecasts 
$P, Q \in \mathcal{P}$, the function $G(P, Q) = \int G(P, y)dQ(y)$ is such 
that $G(Q,Q) \leq G(P,Q)$. If the inequality is strict, then $G(\cdot, \cdot)$ 
is a \textit{strictly proper scoring rule}.

 Proper scoring rules are used not only to score probabilistic forecasts but 
 as functions to optimize over when selecting weights for an optimal linear 
 pool \cite[]{lavine2021adaptive,li2019combining,yao2018using,
 thorey2017online,geweke2011optimal}. 
In practice, a number of scoring rules for continuous distributions are used 
to estimate parameters and evaluate forecasts 
\cite[for some examples]{gneiting2014probabilistic, gneiting2007probabilistic}, 
the LogS being perhaps the most widely used. The LogS is defined in 
(\ref{eq:logs}) and evaluates a distribution by its PDF $p$. 

\begin{equation}
    \label{eq:logs}
    \text{LogS}(p, y) = -\text{log}(p(y))
\end{equation}
The CRPS is an increasingly popular scoring rule which, depending on the 
application, has certain advantages over the LogS. The CRPS is a global 
scoring rule in that it evaluates the whole distribution of a forecast 
whereas the LogS is a local scoring rule in that it evaluates the distribution 
only at a point \cite[]{gneiting2014probabilistic}. When used for estimation, 
the CRPS is often a more robust estimator than the LogS and can lead to 
sharper and better calibrated forecasts under model misspecification 
\cite[]{gebetsberger2018estimation,gneiting2005calibrated}. The CRPS is 
defined in (\ref{eq:crps}) and evaluates a distribution through its CDF $P$.

\begin{equation}
    \label{eq:crps}
    \text{CRPS}(P, y) = \int_{-\infty}^{\infty} (P(x) - \mathds{1} 
    (y \leq x))^2 dx
\end{equation}
For $P$ belonging to certain distribution families, there may be a closed 
form solution to (\ref{eq:crps}) (see table 1 of \cite{zamo2018estimation}). 
Particularly relevant for the work herein is the closed form solution of the 
CRPS when $P$ is a mixture distribution with normal components. 
\cite{li2019combining} provide the derivation for the normal mixture CRPS 
in (\ref{eq:norm_mix_crps}).

\begin{equation}
\begin{aligned}
    \label{eq:norm_mix_crps}
     CRPS(P, y) &= \sum_{c = 1}^C \sum_{c' = 1}^C \alpha_{c, c'}w_c w_{c'} + 
     \sum_{c = 1}^C \beta_c w_c; \quad \text{ where}\\~  \\
        \alpha_{c,c'} = &-\frac{1}{\sqrt{2\pi}}\sqrt{\sigma_c^2 + 
        \sigma_{c'}^2} \text{exp} 
        \left( \frac{(\mu_c - \mu_{c'})^2}{2 (\sigma_c^2 + 
        \sigma_{c'}^2)}\right) \\
        &-\frac{\mu_c - \mu_{c'}}{2}\left[2 \Phi 
        \left(\frac{(\mu_c - \mu_{c'})}{\sqrt{\sigma_c^2 + 
        \sigma_{c'}^2}}\right) - 1 \right] \\
        \beta_c = &\sqrt{\frac{2}{\pi}}\sigma_c 
        \text{exp}\left(-\frac{(\mu_c - y)^2)}{2\sigma_c^2}\right) \\
        &+ (\mu_c - y) \left[2\Phi\left(\frac{\mu_c - y}{\sigma_c}) 
        \right) - 1\right] 
\end{aligned}
\end{equation}

The CRPS has some general forms other than (\ref{eq:crps}) including 
(\ref{eq:crps_mse}) which was shown by \cite{gneiting2007strictly}.

\begin{equation}
    \label{eq:crps_mse}
    \text{CRPS}(P, y) = E|X -y| - \frac{1}{2} E|X - X'|
\end{equation}
Here $X$ and $X'$ are independent copies of the same random variable with 
CDF $P$. When there is no closed form solution to the CRPS but where one 
can sample from the forecast distribution, the CRPS may still be estimated 
via Monte Carlo approximation by estimating the expectations in 
(\ref{eq:crps_mse}). For the cases where $P$ takes the form of an ensemble 
forecast where all components are continuous distributions, we contribute the
additional reduction of
(\ref{eq:crps_mse}) to (\ref{eq:crps_ens_mse}) found in proposition 
\ref{prop:crps_ens_mse}. A 
simple proof for proposition \ref{prop:crps_ens_mse} 
is in appendix \ref{app:crps_ens_pf}. 

\begin{proposition}
Given $C$ CDFs of continuous distributions $P_1, ..., P_C$ and 
a set of weights on a simplex $w_1, ..., w_C$, the CRPS of the mixture
distribution with CDF $\bar{P}(x) = \sum_c w_c P_c (x)$ evaluated at $y$ has
the form of equation (\ref{eq:crps_ens_mse}).
\begin{equation}
    \label{eq:crps_ens_mse}
    CRPS(\bar{P}, y) = \sum_{c = 1}^C w_c E|X_c - y| - 
    \frac{1}{2}\sum_{c = 1}^C \sum_{c' = 1}^C w_c w_{c'} E|X_c - X_{c'}|
\end{equation}
Here $X_c$ is a random variable with corresponding CDF $P_c$, and $X_{c'}$ is 
a random variable independent of $X_c$ but which is equal in distribution.
\label{prop:crps_ens_mse}
\end{proposition}
\cite{li2019combining} showed this result for all mixture components being 
Gaussian, but it's simple to show that it holds when all components are 
continuous. This result allows for straightforward CRPS estimation for a 
continuous mixture distribution forecast when samples of the component 
distributions are available. 
% It's also useful for showing the CRPS is a convex function, an important point for proving theorem \ref{thm:sgp_cons} below (see appendix \ref{app:sgp_cons_pf}).
To complete the definition of the SGP in (\ref{eq:sgp}), we let the function 
$G(\cdot, \cdot) = CRPS(\cdot, \cdot)$ and $G_n$ take the form of either 
(\ref{eq:er_iid}) or (\ref{eq:er_dyn}).

\subsection{SGP consistency}
In \cite{martin2022direct}, parameter asymptotic consistency results of the 
Gibbs posterior similar to consistency results pertaining to the standard 
Bayesian posterior are established, and minimal conditions for consistency 
are given for i.i.d. data. The difference for the Gibbs posterior is that 
the posterior mass converges to the risk minimizer. The conditions for 
consistency follow the conditions for consistency in $M$-estimation 
\cite[]{van1998asymptotic}. We define posterior consistency in 
\ref{def:cons} which is the same definition used by \cite{martin2022direct}.

\begin{definition}
\label{def:cons}
For a given distance $d:\Theta \times \Theta \rightarrow \mathbb{R}^+$, 
the Gibbs posterior distribution $\pi_n^{(\eta)}$ is consistent at 
$\theta^*$ if
\[
    \pi_n^{(\eta)}(\{\theta : d(\theta, \theta^*) > \epsilon\}) \rightarrow 0
    % \text{ in probability, as } n \rightarrow \infty
\]
in probability as $n \rightarrow \infty$
\end{definition}

% \begin{definition}
%   yourmom.
% \end{definition}

For the SGP in (\ref{eq:sgp}) with the empirical risk function in 
(\ref{eq:er_iid}) the posterior consistency is shown for the distance 
function being the Euclidean distance, 
$d(\boldsymbol{a}, \boldsymbol{b}) = ||\boldsymbol{a} - \boldsymbol{b}||$, 
and the prior distribution on the weight vector $\boldsymbol{\omega}$ 
to be the Dirichlet 
distribution $\pi(\boldsymbol{\omega}) \sim Dirichlet(\lambda)$. 
Other distance functions 
and prior distributions may work, but importantly one of the conditions for 
consistency is that the prior distribution gives sufficient mass to the risk 
minimizer or that $\pi(\{\theta:R(\theta) - R(\theta^*) < \delta \}) > 0$ 
for all $\delta > 0$. The support of the Dirichlet distribution is over the 
simplex and thus meets this condition. For i.i.d. data $y_1, ..., y_n$ with 
probability law $\mathcal{L}$ where $E|Y| < \infty$ and where the empirical 
risk function is the function in (\ref{eq:er_iid}), posterior consistency of 
the SGP proposed in (\ref{eq:sgp}) is affirmed by theorem \ref{thm:sgp_cons}. 

\begin{theorem}
\label{thm:sgp_cons}
     
     If $\boldsymbol{\omega}^* \in \Omega = \{w1, ..., w_C : 
     \sum_c w_c = 1, 0 \leq w_c \}$ uniquely minimizes $G(\omega) = 
     E[CRPS(\bar{P}_{\omega}, Y)]$ and for the prior distribution $\pi$, 
     $\pi(\{\boldsymbol{\omega} : G(\boldsymbol{\omega}) - 
     G(\boldsymbol{\omega}^*) < \delta\}) > 0$ 
     for all $\delta > 0$, then for any $\epsilon > 0$

     \[
        \pi_n^{(\eta)} (\{\boldsymbol{\omega} : 
        d(\boldsymbol{\omega}, \boldsymbol{\omega}^*) > \epsilon \} ) 
        \rightarrow 0
     \]
     in $\mathcal{L}$-probability as $n \rightarrow \infty$
\end{theorem}

% \begin{theorem}{\emph{SGP consistency}}

% Let $\{y_i; i = 1, ..., n\}$ be i.i.d. data from a distribution with law $\mathcal{L}$. Let $P$ be a finite mixture distribution of the form of (\ref{eq:}) where each component distribution $P_c$ is continuous and under $P$, $E|X| < \infty$ and $\omega^* \in \{w1, ..., w_C : 0 \leq w_c \leq 1\}$ is the vector of mixture distribution weights which uniquely minimizes $CRPS(\sum_c w_c P_c, Y)$. If $\pi(\omega)$ is a prior distribution assigned to $\omega$ such that 
% \[
%     \pi(\{\omega: S(\omega) - S(\omega^*) < \delta \} ) > 0 \quad \forall \delta > 0,
% \]
% $\pi_n^{(\eta)} (\omega)$ is a Gibbs posterior of the form in (\ref{}) and $d(\cdot, \cdot)$ is the euclidean distance, then for any $\epsilon$

% \[
%     \pi_n^{(\eta)} (\{\omega : d(\omega, \omega^*) > \epsilon \}) \rightarrow 0, \quad \text{ in $\mathcal{L}$-probability as $n \rightarrow \infty$}
% \]

% \end{theorem}

This manuscript does not include consistency results beyond the case of data 
being i.i.d. and having competing models which are fixed, or unaltered by 
additional data. Such is the setting of the first simulation study in the 
next section. One reason it may be difficult to develop theory in a dynamic 
setting is that the models which have the best predictive abilities may be 
changing in time making $\boldsymbol{\omega^*}$ a 
moving target. However, this does not 
necessarily discount the SGP's ability to select model weights which produce 
well performing ensemble forecasts in dynamic settings. This is explored in 
additional simulation studies and in a real data example below.

\section{Simulation studies} \label{sec:simulation}

This section contains three simulation studies to demonstrate the SGP's ability 
to estimate optimal weights for model combination. The first study is done in 
an i.i.d. data setting with fixed competing predictive models. The second study 
includes the same predictive models, but the data is generated dynamically and 
weights are optimized for one step ahead forecasts. The third study focuses on
combining four models used to forecast data generated from the stochastic
compartmental susceptible-infectious-recovered (SIR) disease outbreak model.
Unless stated otherwise, the prior distribution used for the SGP was an
uninformative Dirichlet distribution with the parameter being a vector of 1s.
In all studies, the weights used in the SGP ensembles are selected as the 
mean vectors of 
individual draws from the Gibbs posterior obtained  Hamiltonian 
Monte Carlo sampling using the \texttt{cmdstanr} \texttt{R} package 
\cite[]{stan2024manual,gabry2022stan}.

\subsection{I.I.D. data}

\begin{figure}[hbt!]
    \centering
    \includegraphics[scale = .17]{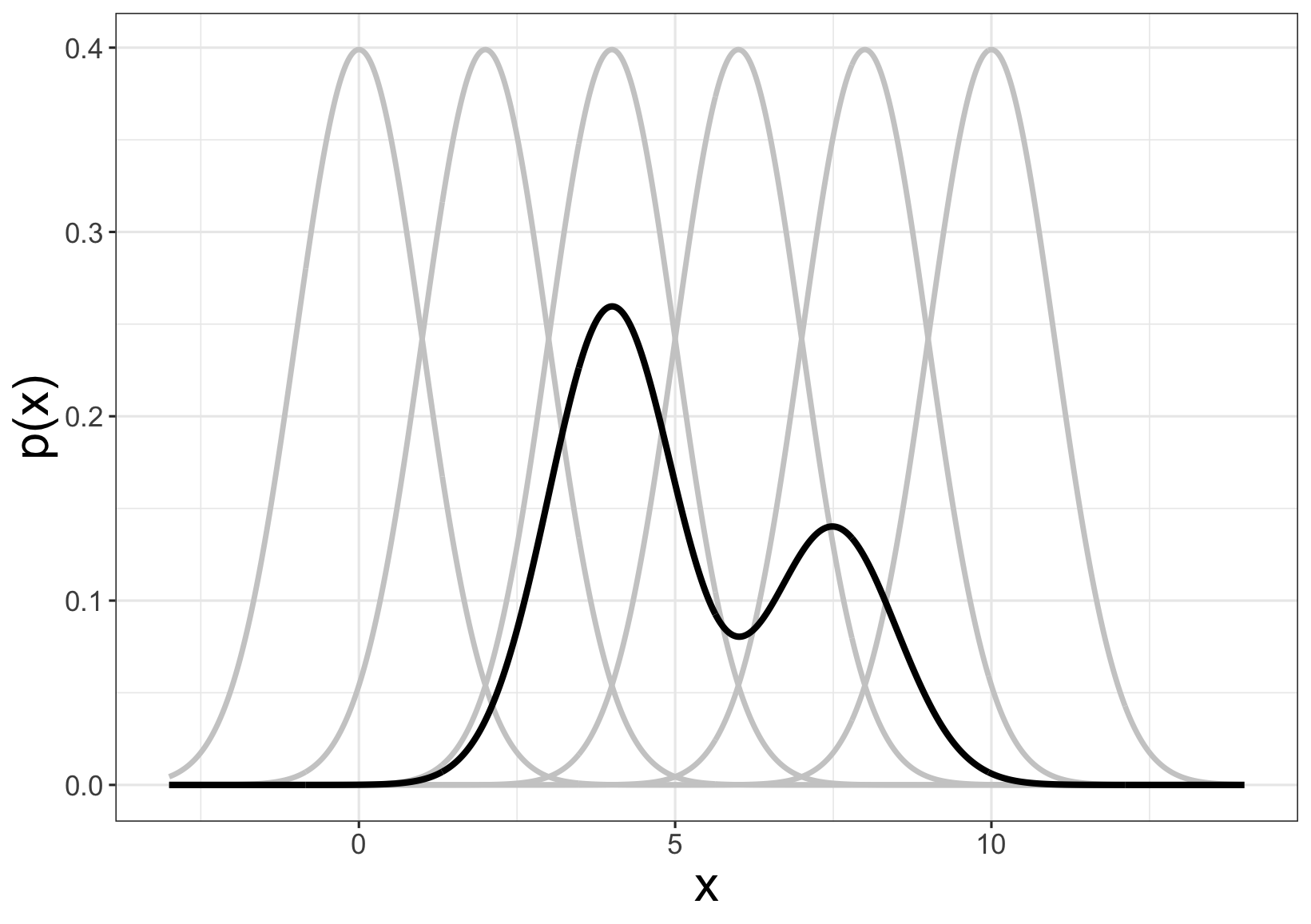}
    \caption{Mixture distribution density function from which data are 
    simulated (black) and 6 candidate normal predictive densities (grey). 
    The mixture distribution is 
    $\nu \times N(3, 1) + (1 - \nu)\times N(6.5,1); \nu = 0.65$. The candidate 
    predictive distributions are each normal with variance 1 and 
    means 0, 2, 4, 6, 8, and 10.}
    \label{fig:iid_sym_examp}
\end{figure}

This first shows how the SGP may be used to optimize 
weights among competing models in an $\mathcal{M}$-open scenario and is similar 
to the first simulation study in \cite{yao2018using}. In this study there are 
six candidate models for predicting the data. Data is generated from the 
mixture of normals distribution 
$\nu \times N(3, 1) + (1 - \nu) \times N(6.5, 1)$ where $\nu = 0.65$. 
Because all simulated data is i.i.d., the empirical risk function from 
(\ref{eq:risk_min}) is used, and the consistency results from theorem 
\ref{thm:sgp_cons} hold. The six competing models are each normally 
distributed with variance 1, but each model has a unique mean parameter, 
the unique means being $\{0, 2, 4, 6, 8, 10\}$. Figure \ref{fig:iid_sym_examp}
% \spencer{(may put in supplementary materials)}
shows in black the mixture distribution from which data is simulated as well as 
the candidate normal predictive models in grey.

The predictive performance of the SGP ensemble 
is compared with that of BMA, AVS, and EQW. For 
both AVS and SGP, the learning rate $\eta$ needed to be selected, and we opted 
to select $\eta$ with the aim of minimizing the CRPS for prediction. A figure
in the supplementary materials shows an example of how the 
CRPS varies for different 
values of $\eta$ for the AVS and SGP. In the case of AVS, LOO cross 
validation over a grid of values for $\eta$ was performed and the $\eta$ 
which minimized the predictive CRPS was selected. For the SGP, the value of 
the CRPS continues to decrease as $\eta$ increases. The improvement became 
less stark for higher values, so $\eta$ was fixed at 15 so as to retain 
influence from the prior distribution.

\begin{figure}[hbt!]
    \centering
    \includegraphics[scale=.2]{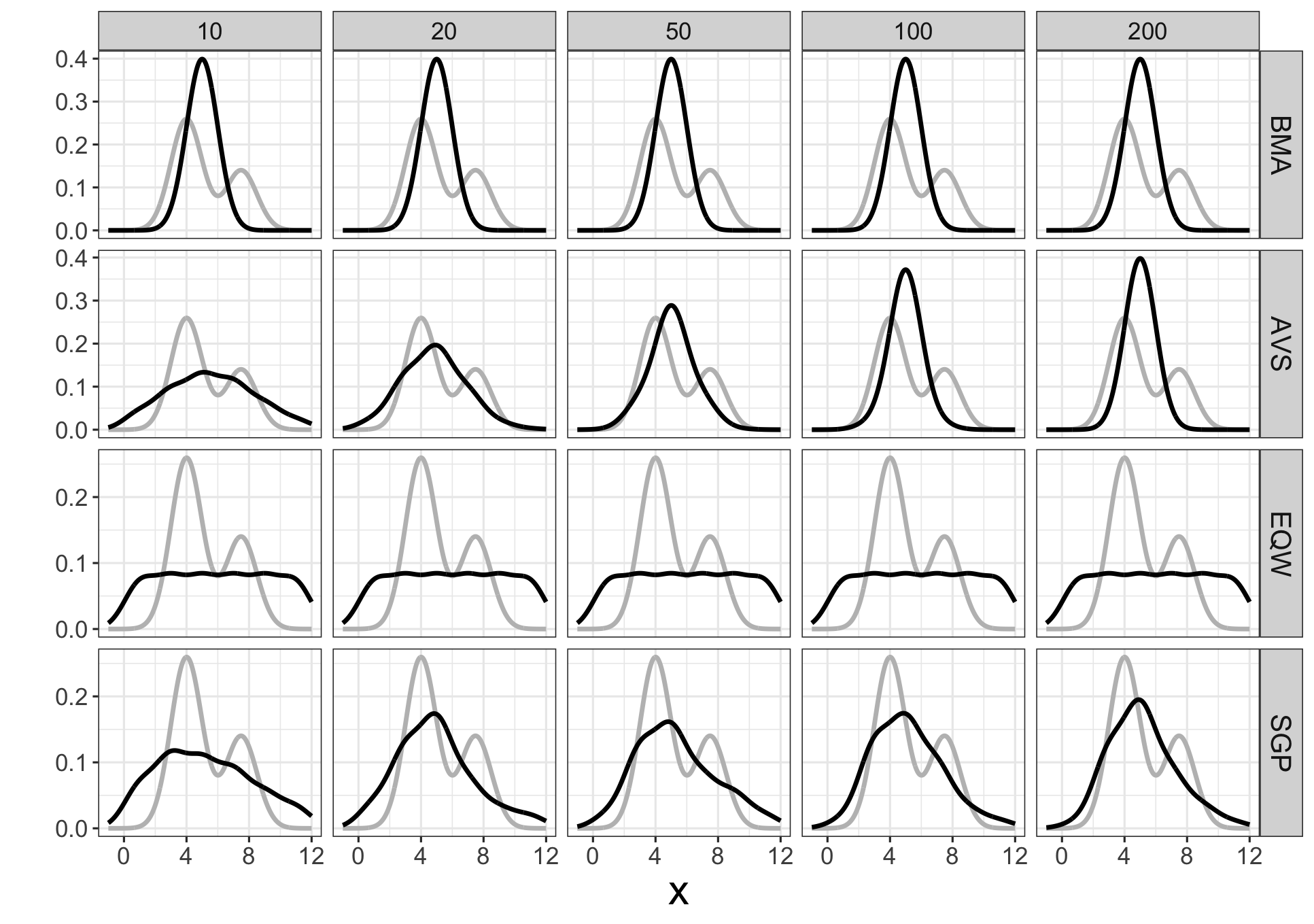}
    \caption{Examples of estimated densities after weighting of competing 
    models faceted by weighting method and sample size. 
    The weighting methods shown include BMA, AVS, EQW, and SGP. The estimated 
    densities (black) overlay the true model density (grey).}
    \label{fig:opt_dens}
\end{figure}

Figure \ref{fig:opt_dens} %(\spencer{put in supplement?})
shows examples from estimating model weights of 
constructing a linear pool from the candidate models for samples sizes of 10, 
20, 50, 100, and 200 for BMA, AVS, EQW, and SGP weight selection methods. 
The top 
row shows how with a sample size as small as 10, BMA concentrates nearly all 
model probability into one model. AVS does better at spreading the model 
probability around, but as the sample size increases, it too will often favor 
one model over all others. SGP on the other hand tends to better capture the 
spread of the true distribution as the sample size increases.

\begin{figure}[hbt!]
\centering
  \includegraphics[width=.98\linewidth]{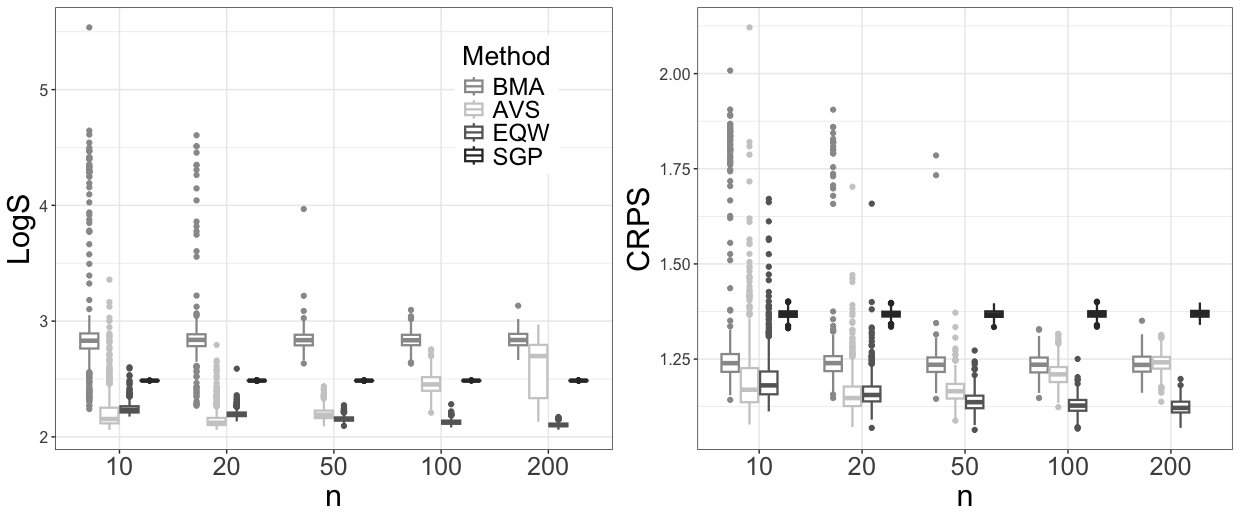}
\caption{Boxplots of the mean of 1,000 Monte Carlo LogSs (left) and CRPSs 
(right) for 500 replicates and for increasing sample size $n$. 
Plots are colored by weighting method. For both
LogS and CRPS, lower is better. }
\label{fig:iid_scores}
\end{figure}

Figure \ref{fig:iid_scores} shows boxplots of the CRPS and the LogS for the 
four competing weight selection methods. 
The simulation was replicated 500 times for 
each sample size, and after selecting weights, the CRPS and LogS in these plots 
were calculated as the Monte Carlo mean scores for 1,000 additional draws from 
the true distribution after selecting weights. As the sample size increases, 
the SGP clearly separates itself from the model averaging approaches. AVS 
appears to perform well for smaller sample sizes but worsens as the sample 
size increases, and it tends to favor one model over the others.

\jarad{Include "lower is better".}
\jarad{Write out CRPS and LogS}
\jarad{Make n explicit}

\subsection{Dynamic data simulation forecast}

The simulation study in this section has similarities to the previous study 
including that it utilizes the same fixed candidate predictive models and the 
data is simulated from a normal mixture distribution. Here, however, the 
mixture distribution from which the data are simulated is dynamic over time 
and the focus of making an ensemble model is to perform one step ahead 
forecasting. 
The data were simulated from the model in (\ref{eq:dyn_sim_mod}). The two 
component normals in the mixture distribution are the same as in the previous 
study, but here the weights for each model are updated at each time point 
according to a random walk process. 
\begin{equation}
    \begin{aligned}
    \label{eq:dyn_sim_mod}
        Y_t &\sim \omega_{1,t} \times N(3, 1) + \omega_{2,t} 
        \times N(6.5, 1) \\
        \omega_{i,t} &= \frac{\text{exp}(z_{i,t})}{\sum_a \text{exp}(z_{i,t})}; 
        \quad i \in \{1,2\}\\
        \boldsymbol{z}_t &\sim N(\boldsymbol{z}_{t-1}, \sigma^2 I) \\
        (\omega_{1,1}, \omega_{2,1}) &= (0.65, 0.35) \\
        \sigma^2 &= 0.01
    \end{aligned}
\end{equation}
A single observation $y_t$ is simulated at each time step 
$t \in \{1, ..., T = 50\}$, and at each step the ensemble weights are 
selected and an ensemble for forecasting $\tilde{y}_{t+1}$ is constructed. 
A figure 
% \ref{fig:dyn_wts_ex} 
in the supplementary materials
gives an example of the dynamic weights simulated 
from (\ref{eq:dyn_sim_mod}). The left plot shows the simulated weight values, 
and the right plot shows the SGP optimized weights of the competing forecasts 
for times 2 to 50. The empirical risk used for fitting the SGP is that in 
(\ref{eq:er_dyn}). The same discount factor of $\alpha^{T-t}$
is similarly applied to the AVS and BMA methods as done in 
\cite{lavine2021adaptive}.

\begin{figure}[hbt!]
\centering
  \includegraphics[width=.8\linewidth]{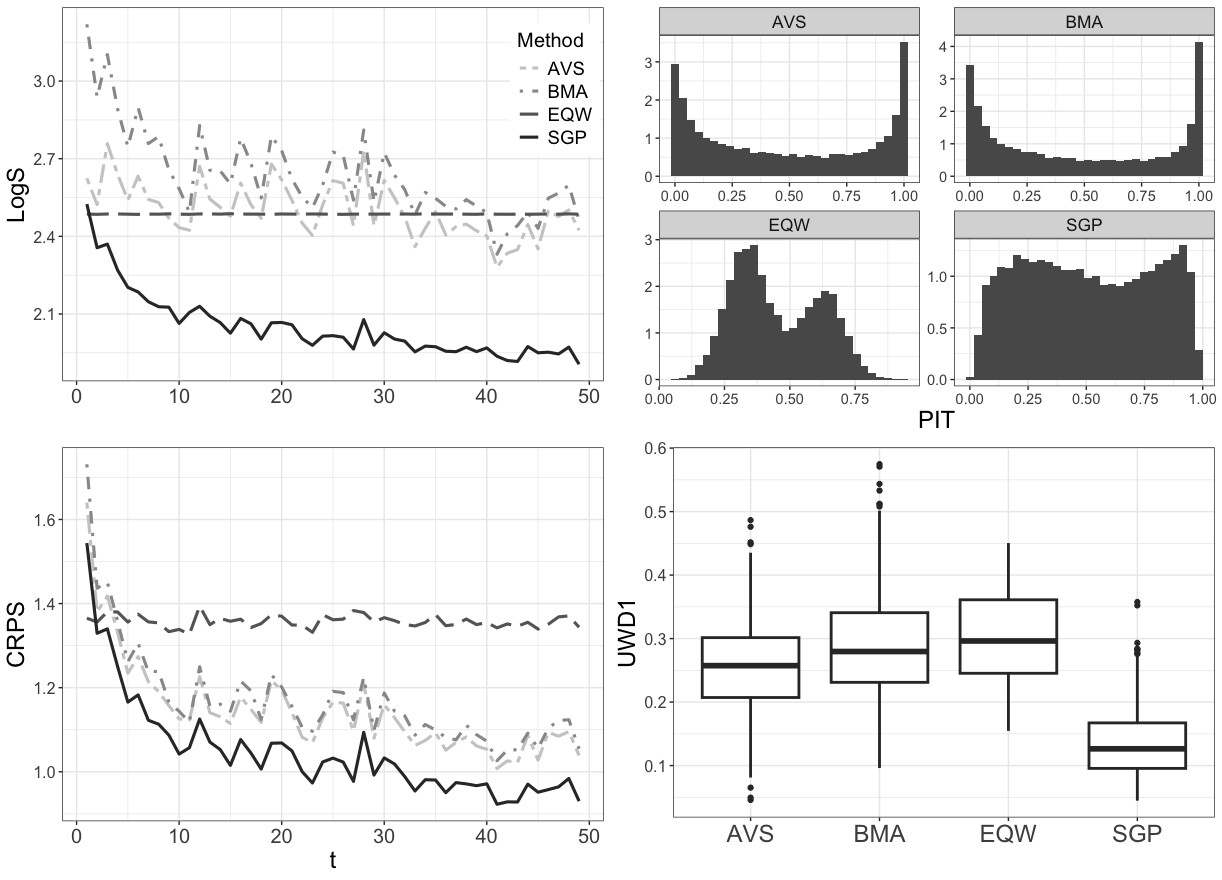}
\caption{Plots showing results for one step ahead forecasts for dynamically
weighted data for the four
weighting methods SGP, AVS, BMA, and EQW. Two plots show performance
according to proper scoring rules
(left). The mean LogS (top) and CRPS (bottom) over 500 replicates of one step
ahead predictions at 49 time points colored by weighting methods.
The other two plots show calibration according to the PIT (right). The PIT
histogram for all forecasts faceted by weighting method is shown (top), and
boxplots by weighting method of the UWD1 for all forecasts of one simulation
replicate is shown (bottom).}
\label{fig:dyn_sim}
\end{figure}

The performance of the ensemble forecasts is assessed in terms of proper 
scoring rules as well as by the PIT, that is the uniformity of the histogram 
of observations evaluated by the probabilistic forecast. The PIT is evaluated 
both by looking at the PIT histogram and by measuring the distance between the 
PIT and the uniform distribution using the unit Wasserstein distance metric 
(UWD1) from section 3.4 in \cite{wadsworth2025quantile}.
The plots on the left of figure \ref{fig:dyn_sim} show how the CRPS and LogS 
forecasts behave over time for one step ahead forecasts. 
The lines represent the mean scores over the 500 
simulation replicates. Here the SGP outperforms AVS, BMA, and EQW ensembles 
with the superior performance appearing more starkly for LogS than for CRPS. 
Unsurprisingly the EQW forecast scores remain relatively constant through time 
while the other methods improve as time goes on. Interestingly, the behavior 
of AVS shown in the i.i.d. case in the previous study to perform very well 
with small amounts of data but to worsen with larger data does not manifest 
in this study of dynamic forecasting.
The right two plots in figure \ref{fig:dyn_sim} show 
the PIT results for one step ahead forecasts for forecasts of times 2 to 50 
over the same 500 replications of simulated data. The histogram plots clearly 
show that the
SGP PIT histogram is much nearer to uniformity than the PIT histograms for the 
other methods. The boxplots of the 500 UWD1 distances show smaller UWD1 values 
for the SGP PIT meaning the distances between the PIT and the standard uniform 
distribution tend to be smaller for SGP, further showing superior calibration 
compared to the other methods.

% \begin{figure}[hbt!]
% \centering
% %\begin{subfigure}{}
%   % \centering
%   \includegraphics[width=.48\linewidth]{Images/dyn_sim_pit.png}
%   % \caption{A subfigure}
%   % \label{fig:sub1}
% %\end{subfigure}%
% \hspace{0.01\textwidth}
% %\begin{subfigure}{}
%   \centering
%   \includegraphics[width=.48\linewidth]{Images/uwd1_box.png}
%   % \caption{A subfigure}
%   % \label{fig:sub2}
% %\end{subfigure}
% \caption{Plots for assessing calibration of one step ahead forecasts for the four weighting methods. The figure shows the PIT for all $99 \times 500$ simulated one-step-ahead forecasts for AVS, BMA, EQW, and SGP (left) and boxplots of the UWD1 distance between a standard uniform distribution and the empirical distribution of the PIT for the weighting schemes for 99 predictions (right).}
% \label{fig:dyn_pit}
% \end{figure}

\subsection{SIR data simulation forecasts}

The SIR compartmental model was introduced by \cite{kermack1927contribution} 
and is a standard mathematical model used for modeling
disease outbreaks. The SIR model models a population in which at any given 
time each individual belongs to one of
three compartments: susceptible, infected, or recovered. The most basic SIR
model is deterministic and defined by a set of ordinary differential equations
--one for each compartment-- but stochastic extensions allow one to simulate
from an SIR model with uncertainty 
\cite[]{allen2008mathematical,widgren2019siminf}. When fitting an SIR model
to track an epidemic, it is common practice to treat real outbreak measurements
as analogous to the values in the infected compartment
\cite[]{widgren2019siminf, osthus2019dynamic}. 
% \spencer{Should I add more details
% about the SIR model, maybe in the supplementary material?}

The data used in this simulation study were generated from a stochastic SIR 
model using the \texttt{SimInf} package in \texttt{R} 
\cite[]{widgren2019siminf}. For each of 3,000 simulation replicates, 35 
consecutive ``weeks`` of infection counts were sampled. Starting at week 5, 
we fit four individual forecast models and used each to make one week ahead 
forecasts of 
infections for each of weeks 6 to 35.
Using the individual forecasts, we used the SGP to construct an 
ensemble forecast and compared the performance to the performances of 
ensembles constructed via the AVS, BMA, and EQW methods. The decision to use
four component forecasts was based on the recommendation from 
\cite{fox2024optimizing} to forecasts diseases such as influenza and COVID-19
with an ensemble of four or more components. The four models included 
a Bayesian SIR model similar to that used to forecast the flu in 
\cite{osthus2019dynamic}, a nonlinear Bayesian model similar to that used by 
\cite{ulloa2019} and \cite{wadsworth2025forecasting}, an ARIMA model fit 
using the \texttt{forecast} package in \texttt{R} 
\cite[]{hyndman2008automatic}, and a naive random walk model also fit using
the \texttt{forecast} package. The two Bayesian models were fit using the 
proportion of infections of the total population. 

\begin{figure}[hbt!]
\centering
  \includegraphics[width=.8\linewidth]{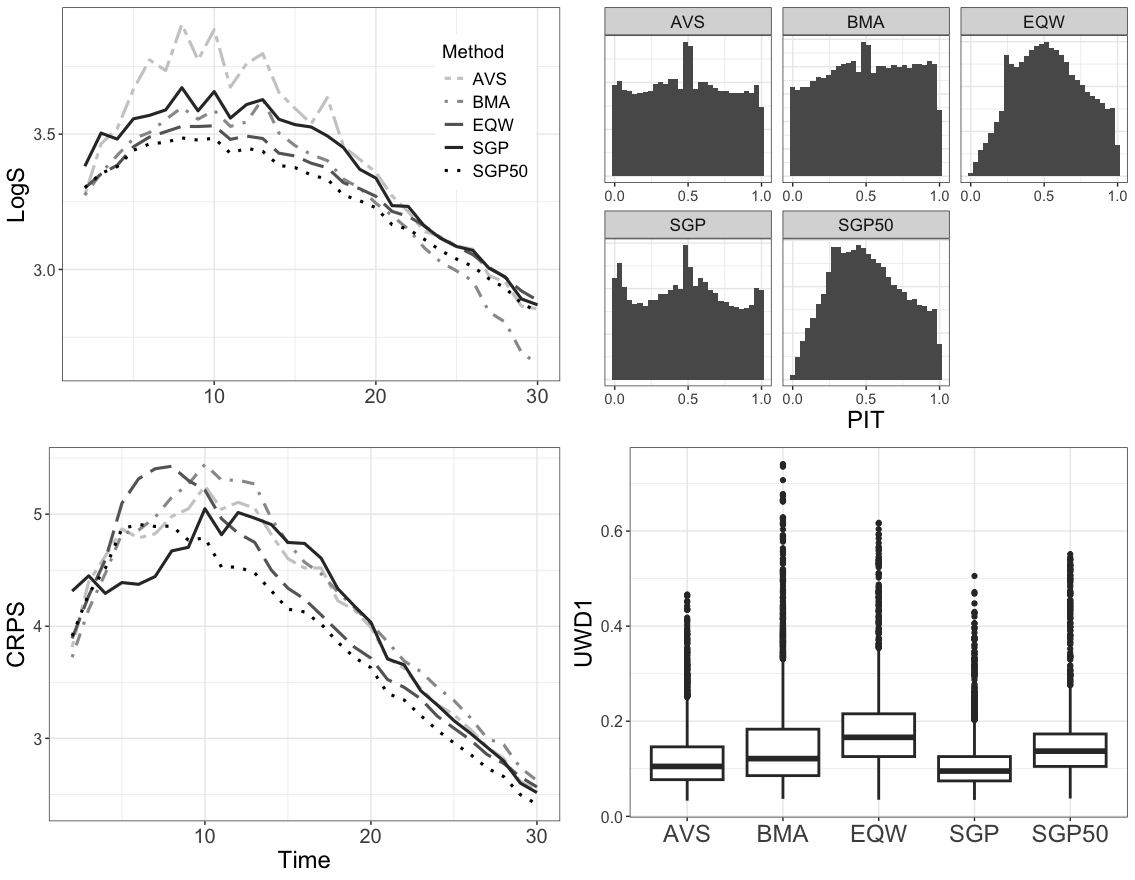}
\caption{Plots showing results for one step ahead forecasts for SIR
data for the five
weighting methods SGP, SGP50, AVS, BMA, and EQW. Two plots show performance
according to proper scoring rules
(left). The mean LogS (top) and CRPS (bottom) over 3,000 replicates of one step
ahead predictions at 29 time points colored by weighting methods.
The other two plots show calibration according to the PIT (right). The PIT
histogram for all forecasts faceted by weighting method is shown (top), and
boxplots by weighting method of the UWD1 for all forecasts of one simulation
replicate is shown (bottom).}
\label{fig:sir_sim}
\end{figure}

All forecasts were given
as 10,000 draws from a predictive distribution, and using the draws, the 
expectation components in \ref{eq:crps_ens_mse} were estimated in order to 
evaluate the CRPS and select component weights. We constructed
two ensembles using the SGP, one using a less informative prior distribution
and one using a more informative prior distribution. The prior distributions
were both uniform Dirichlets, that is Dirichlet distributions with equal
expected weights for each component, but with differing tightness around 
the prior expectation. For one the prior parameter vector was
$\mathbf{1}_4$, and for the other the prior parameter vector was 
$\mathbf{50}_4$. The latter distribution assigns a tighter probability to the
equal weights and a thus a stronger regularization to an equally weighted 
ensemble. In figure \ref{fig:sir_sim}, these two ensembles are denoted as
SGP and SGP50 respectively. Unlike in the previous study, the value of 
the learning rate $\eta$
did not appear to affect forecast performance, so for both AVS and SGP methods
we fixed $\eta = 1$.

Figure \ref{fig:sir_sim} shows the results of the one step ahead forecasts for
3,000 simulation replicates. The lines in the LogS and CRPS plot the
average forecast scores over the 3,000 replicate by method. Here the
SGP ensemble does not perform as well as in the previous two studies when
compared with other methods, but 
the SGP50 with the stronger regularization prior performs very well, resulting
in lower
LogS and CRPS scores than the other methods for most of the 30 weeks shown. 
A figure 
% \ref{fig:sir_sim_med} 
in the supplementary material shows similar plots
for the forecasts scores differing only in that the lines represent the 
median of scores
rather than the mean. There the SGP and AVS ensembles appear to perform the
best of all methods. The results suggests that the regularization to 
equally weighted 
forecast components makes ensembles forecasts robust to particularly high 
scores.
The PIT histograms and boxplots in figure \ref{fig:sir_sim}
show that relative to the other methods, the SGP performs well in terms of 
forecast calibration. The SGP50 is less well calibrated and the histogram has a 
shape much closer to that of the EQW ensemble forecasts.

The studies in this section show the SGP's ability to construct accurate 
forecasts for cases of both i.i.d. and dynamic data. The SGP largely 
outperformed AVS, 
BMA, and EQW ensembles both in terms of mean proper scoring rules and in PIT 
uniformity and calibration. 
The value of regularization to equal ensemble weights using a 
stronger prior distribution was also shown in the third study. 
In the next section, the SGP ensemble construction is applied 
to forecasts from the CDC FluSight, again with comparisons of 
performance against AVS, BMA, and EQW.

\section{Analysis on 2023-24 CDC flu forecast competition} \label{sec:flu_analysis}

Every flu season, beginning in 2013 but with the exception of the 2020-21 flu 
season, the CDC has hosted a national flu forecasting competition, also known 
as FluSight. The competition begins in October and lasts around 30 weeks until
May of the following year. The competition generally involves 
a few dozen academic and 
industry research teams who each create separate probabilistic forecasts 
targeting certain flu events for future weeks. At the end of the season, 
forecast performance of all forecasts from each team are evaluated, and a 
winning team is announced 
\cite[]{mathis2024evaluation,biggerstaff2016results}. During the 2022-23 
and 2023-24 flu seasons, forecasts targeted 1, 2, 3, and 4-week ahead 
hospitalizations due to a laboratory confirmed flu infection for each of the 
50 US states, District of Columbia, Puerto Rico, and the nation as a whole. 
Every week during the competition, a team could submit a forecast for each 
target, and after the hospitalization count for the target was observed, the 
forecast was then scored and compared with all other forecasts. Besides 
reporting and scoring all forecasts individually, all submitted forecasts 
were combined into an ensemble forecast, which ensemble forecast 
was then published at 
\cite{cdc2024previous} as the official CDC forecast and scored along 
with the individual forecasts 
\cite[]{mathis2024evaluation,wadsworth2023mixture}.  

Teams were given complete freedom in how they developed their forecasts, 
and since the beginning of the competition there has been a large variety of 
methodologies used 
\cite[]{mathis2024evaluation,mathis2023flusight,mcandrew2021adaptively,
osthus2021multiscale,osthus2019dynamic,ulloa2019,morgan2018wisdom,
farrow2017human}. To make scoring and ensemble construction straightforward, 
each submitted forecast was required to follow a specified representation. 
During the 2022-23 and 2023-24 seasons, the requested forecast representation 
was a set of estimated quantiles for each of 23 probability values established 
by FluSight. For notation, we say each forecast team is given $K$ 
probabilities $0 \leq p_1, p_2, ..., p_K \leq 1$. Then, the $c^{th}$ 
competing team would submit quantiles --estimated however they please-- 
$q^{(c)}_{r,w+h,1}, ..., q^{(c)}_{r,w+h,K}$ where 
\[
    P(H_{r,w+h} < q^{(c)}_{r,w+h, k}) = p_k; \quad k \in (1, 2, ..., K)
\]
for $H_{r,w+h}$ hospitalizations in location or region $r$, for time horizon 
$h \in \{1,2,3,4\}$ weeks ahead, and where $H_{r,w}$ is the most recently 
observed hospitalization count at week $w$. 

The forecasts were scored by the weighted interval score (WIS), defined in 
(\ref{eq:wis}) \cite[]{bracher2021evaluating}. The WIS is a proper scoring 
rule made for scoring quantile or interval representation forecasts. In the 
definition, $P$ represents a probabilistic forecast with an interval 
representation including $I$ predictive intervals with different nominal 
levels. The values $v_0 = 1/2$ and $v_i = \alpha_i / 2$ are weights for 
each interval where $\alpha_i$ is the nominal level of the $i^{th}$ interval, 
$IS_{\alpha}$ is the interval score (IS) a proper scoring rule for a single 
interval as defined in (\ref{eq:is}), and $y^*$ is the true observation 
targeted by $P$.

\begin{equation}
\label{eq:wis}
        WIS_{0,I}(P, y^*) = \frac{1}{I + 1/2} \times 
        (v_0\times |y^* - median| + \sum_{k=1}^K \{v_i 
        \times IS_{\alpha_i}(P, y^*) \} )
\end{equation}

\begin{equation}
\label{eq:is}
        IS_{\alpha}(l,r;y^*) = (r-l) + 
        \frac{2}{\alpha}(l - y^*)\mathds{1}\{y^* < l\} + 
        \frac{2}{\alpha}(y^* - r) \mathds{1}\{y^* > r\}
\end{equation}
Like the CRPS, the WIS is a global scoring rule, allowing for evaluation over 
the whole range of the forecast. In fact, as the number of forecast intervals 
$I$ increases, the WIS becomes arbitrarily close to the CRPS 
\cite[]{bracher2021evaluating,gneiting2011comparing}. 

A drawback of the quantile representation and the WIS is that no information 
about the forecast distribution exists below the $1^{st}$ quantile or above 
the $K^{th}$ quantile. Another drawback is that even with a large number of 
quantiles, the information is less than would available from a PDF or CDF, 
and the quantile representation is not suited for scoring by the LogS or CRPS 
or ensembling building using a linear pool \cite[]{wadsworth2023mixture}.  
In order to apply the SGP from (\ref{eq:sgp}), which is based on the CRPS, 
the quantile 
Gaussian process quantile matching model from
\cite{wadsworth2025quantile} was fit to the quantiles to approximate a 
continuous
distribution from which the quantiles were estimated. This was done via 
Bayesian posterior Monte Carlo sampling from a normal mixture distribution. The 
result was 50,000 posterior predictive draws for forecasts of each location, 
week, and horizon. Monte Carlo approximations of the expected values in 
(\ref{eq:crps_ens_mse}) were calculated from the posterior predictive samples 
allowing for estimation of the CRPS for individual forecasts and ensemble 
forecasts. This allowed for using the SGP for selecting weights and 
constructing an ensemble forecast.

The analysis of ensemble forecasts of flu hospitalizations herein was limited 
to 1-week ahead forecasts. For each location, only the forecasts from teams 
who submitted forecasts for every week during FluSight were included 
so as to avoid dealing with missing forecasts. We expected that 
competing forecasts would vary in skill throughout the flu season and thus 
used the empirical risk function from (\ref{eq:er_dyn}) used in the dynamic 
setting. The FluSight candidate forecasts were dynamic, changing every week 
as the individual 
forecast teams updated their forecasts and the flu season progressed.
To complete the SGP, an uninformative Dirichlet prior with parameter 
$\mathbf{1}_C$ was assigned to $\boldsymbol{\omega}$. 
The SGP was fit via Hamiltonian 
Monte Carlo sampling using the \texttt{cmdstanr} \texttt{R} package 
\cite[]{stan2024manual,gabry2022stan}. To assess convergence of the fit, 
the SGP for US forecasts at the national level for the week of 
January 20, 2024 (week 15 of 29 of the flu season) was fit with 4 chains. 
Each chain consisted of 60,000 posterior draws, the first 10,000 draws being 
discarded as a burn-in. From this fit, among all posterior parameters, the 
largest $\hat{R}$ statistic \cite[]{vehtari2021rank} was 1.000055, giving 
no reason to be concerned about about MCMC convergence. The smallest 
effective sample size \cite[]{gelman2013bayesian} over all chains was 74,167. 
The SGPs for all other regions and weeks were fit using one chain of 60,000 
draws, the first 10,000 being discarded as a burn-in. The result was 50,000 
Monte Carlo vector draws for the weights $\{\omega^{(m)}\}_{m = 1}^M$.
For each $c$, the Monte Carlo mean of each weight 
$\bar{w}_{c,t} = M^{-1}w^{(m)}_{c,t}$ was calculated, and the mean weights 
were used in the ensemble forecast for hospitalizations at time $t+1$. 
Selecting the learning rate $\eta$ via cross validation did not appear to 
improve forecasts,
so $\eta$ was set to equal 1.
% selected by fitting the SGP for all forecasts up 
% to time $t-1$ and estimating model combination weights at all values of 
% $\eta$ on a pre-selected grid. The value of $\eta$ which produced combination 
% weights leading to the smallest CRPS of an ensemble model at time $t$ was 
% used in the SGP for time $t+1$.
Using this procedure, 1-week ahead ensemble forecasts were constructed for 
flu season weeks 2-29 for each location, noting that for the week 1 forecast 
ensemble, model weights were all selected to be equal. The CRPS was 
evaluated at the log forecast and log hospitalization. The mean CRPS for 
region $r$ across the season was calculated as
\[
    \overline{CRPS}(\bar{P}, \boldsymbol{H}_r) = 
    \frac{1}{W}\sum_{w = 1}^{W-1} CRPS(\bar{P}_w, H_{r,w + 1})
\]
for each region. A similar mean for each week where the average was over 
all regions was also calculated.

% \begin{figure}[hbt!]
%     \centering
%     \includegraphics[scale=.5]{Images/us_post_int.png}
%     \caption{90\% credible intervals for forecast weights estimated via SGP 
%     for the 2023-24 CDC FluSite targetting flu hospitalizations at the 
%     national level of the United States. The intervals in the bottom right 
%     corner are those from the uninformative Dirichlet prior.}
%     \label{fig:us_wt_post_int}
% \end{figure}

\begin{figure}[hbt!]
\centering
%\begin{subfigure}{}
  % \centering
  \includegraphics[width=.48\linewidth]{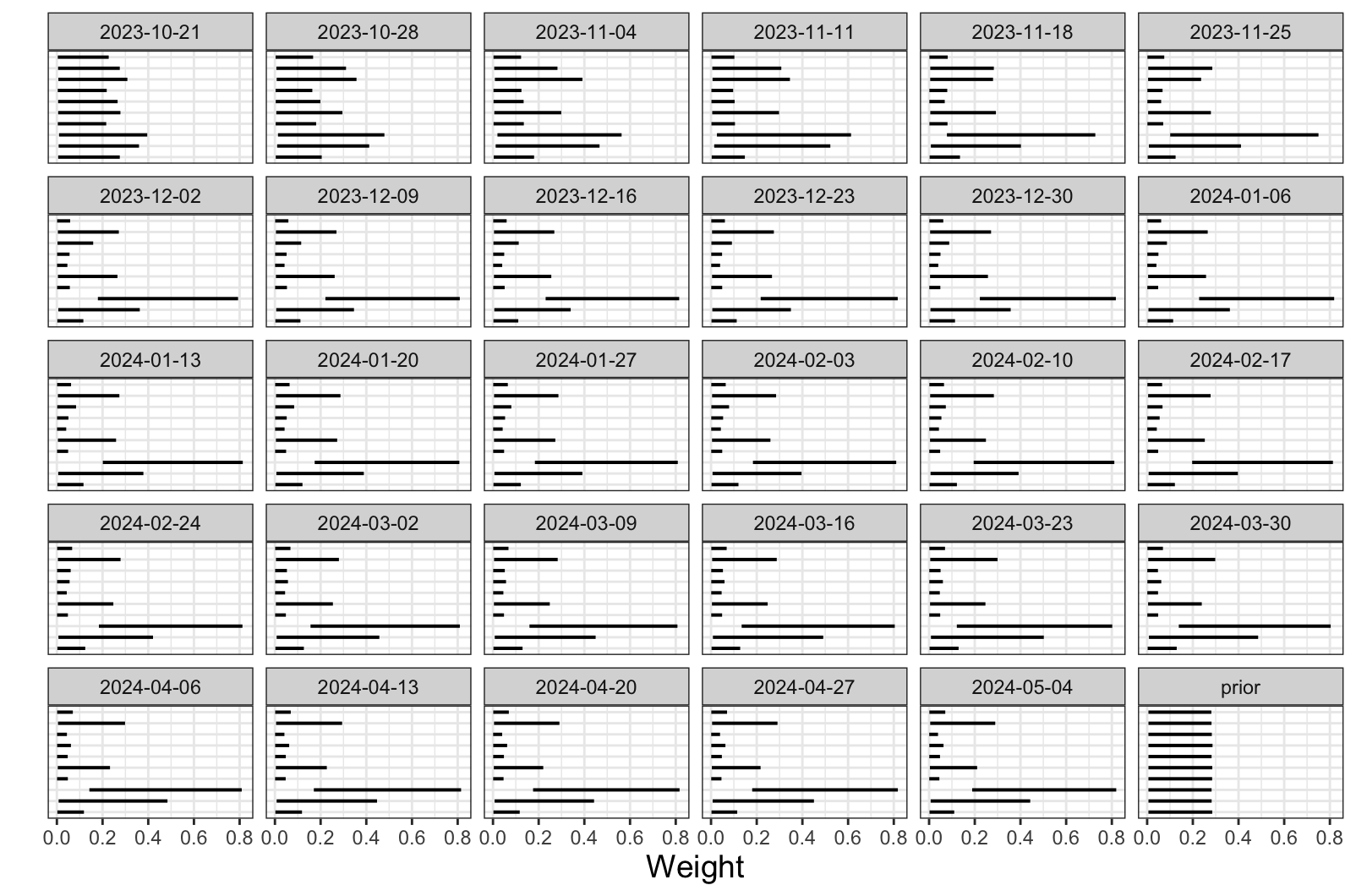}
  % \caption{A subfigure}
  % \label{fig:sub1}
%\end{subfigure}%
\hspace{0.01\textwidth}
%\begin{subfigure}{}
  \centering
  \includegraphics[width=.48\linewidth]{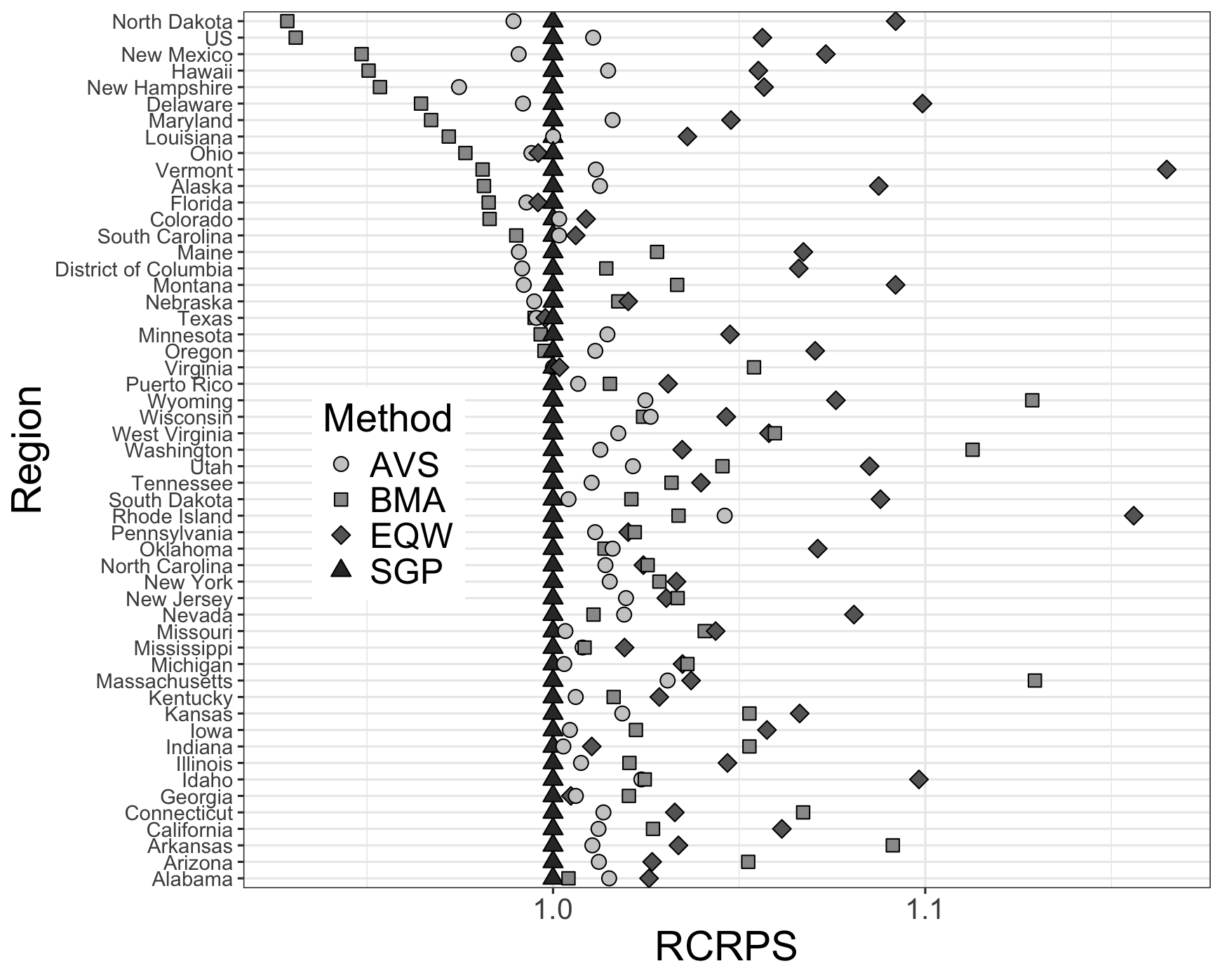}
  % \caption{A subfigure}
  % \label{fig:sub2}
%\end{subfigure}
\caption{90\% credible intervals for forecast weights estimated via SGP 
    for the 2023-24 CDC FluSite targetting flu hospitalizations at the
    national level of the United States (left). The intervals in the bottom 
    right
    corner are those from the uninformative Dirichlet prior.
Mean over regions CRPS for the four ensemble methods divided
by mean CRPS for the SGP
separated by shape and color (right).}
\label{fig:flu_crps}
\end{figure}

The plots on the left of figure \ref{fig:flu_crps} shows 90\% SGP 
credible intervals for the 10 
included forecast models for flu hospitalizations for the state of 
Rhode Island --one of the regions where the SGP forecasts peprformed the best
relative to other methods--
for all weeks during the 2023-24 flu season. The plots show how the SGP model 
probabilities change through the season. At the beginning of the season, there
was little separation in probabilities for individual forecast weights, but
as the season progresses several component forecasts decrease in importance
while a few increase but with higher variability.
The plot on the right of figure \ref{fig:flu_crps} shows the mean CRPS 
scores for 
SGP, AVS, BMA and EQW divided by the mean CRPS score of the SGP
for all regions in one plot and over all weeks in the other. The value for 
SGP is thus always 1, and any value less than 1 for any other method 
is an instance of an ensemble 
method outperforming the SGP for one region.
In 31 
locations, the SGP ensemble forecasts has the best overall scores, and in
only 3 regions does the SGP perform the worst among the methods compared.

Table \ref{tab:forecast_rank} summarises the forecast performance for the 
four methods by showing in how many instances forecasts for each method were 
ranked 1, 2, 3, or 4 by mean CRPS over regions or weeks, a smaller 
ranking meaning better performance. 
By these rankings, SGP performed the best followed by BMA, AVS, and finally
EQW. 
In the supplementary material, a simlilar figure and table to 
% Figure \ref{fig:flu_wis} and table \ref{tab:forecast_rank_wis} are
% similar to those of 
figure \ref{fig:flu_crps} and table
\ref{tab:forecast_rank} show results for ensembles constucted via 
quantile averaging and forecasts scored by the WIS instead of the CRPS.

\begin{table}[h!]
\centering
\caption{Table showing the number of times ensemble forecasts for each method, 
SGP, AVS, BMA, and EQW, were ranked 1st, 2nd, 3rd, and 4th according to mean 
CRPS by region or week. There were 53
regions evaluated and 29 weeks, so each column under region should sum to 53
and each column under week to 29.}
 \begin{tabular}{cccccccccc}
 % {@{\extracolsep{\fill}}
 %    l
 %    *{8}{c}
 %    S[table-format=4.0]}
  Ranking
  & \multicolumn{4}{c}{By region CRPS}
  &
  & \multicolumn{4}{c}{By week CRPS} \\
  \cmidrule{2-5}  \cmidrule{7-10}
  & AVS & BMA & EQW & SGP & & AVS & BMA & EQW & SGP\\
  \midrule
  $1^{st}$ & 5 & 17 & 0  & \textbf{31} & & 5  & 8  & 2  & \textbf{14}\\
  $2^{nd}$ & 32 & 5 & 1  & 15 & & 17 & 4  & 0  & 8\\
  $3^{rd}$ & 16 & 17& 16 & 4  & & 7  & 10 & 8  & 3\\
  $4^{th}$ & 0 & 14 & 36 & 3  & & 0  & 7  & 19 & 3\\
  \bottomrule
 \end{tabular}
 \label{tab:forecast_rank}
\end{table}

% \begin{table}
% \caption{Table showing the number of times ensemble forecasts for each method, 
% SGP, AVS, BMA, and EQW, were ranked 1st, 2nd, 3rd, and 4th. There were 53 
% regions evaluated and 29 weeks, so each column under region should sum to 53 
% and each column under week to 29.}
% \begin{tabular*}{ccccccccc}
% {@{\extracolsep{\fill}}
%     l
%     *{8}{c}
%     S[table-format=4.0]}
%   Ranking
%   & \multicolumn{4}{c}{By region CRPS}
%   & \multicolumn{4}{c}{By week CRPS} \\
%   \cmidrule{2-5} \cmidrule{6-9}
%   & AVS & BMA & EQW & SGP & AVS & BMA & EQW & SGP\\
%   \midrule
%   $1^{st}$ & 0 & 0 & 0 & 53 & 1 & 0 & 0 & 28\\
%   $2^{nd}$ & 44 & 3 & 6 & 0 & 23 & 1 & 5 & 0\\
%   $3^{rd}$ & 7 & 3 & 43 & 0 & 5 & 0 & 24 & 0\\
%   $4^{th}$ & 2 & 47 & 4 & 0 & 0 & 28 & 0 & 1\\
%   \bottomrule
% \end{tabular*}
% \label{tab:forecast_rank}
% \end{table}

% \begin{figure}[hbt!]
% 
%     \centering
%     \includegraphics[scale=.45]{Images/flu_region_crps.png}
%     \caption{Average CRPS for linear pooled forecast across all weeks of the 2023-24 influenza season for all 53 regions. Stacking weighting methods are separated by shape/colour.}
%     \label{fig:opt_dens}
% \end{figure}

This analysis demonstrates the effectiveness of selecting weights for a 
linear pooled ensemble forecast by optimizing the dynamic risk function in 
(\ref{eq:er_dyn}) via the SGP. Given that the forecasts 
used were posterior predictive samples from a model fit on quantile 
forecasts, it was key to be able to make a Monte Carlo approximation of the 
CRPS from (\ref{eq:crps_ens_mse}). Of the methods compared, the SGP is 
arguably the most complicated, but in forecast performance it more often than 
not outperformed the other methods.

\section{Discussion} \label{sec:4conc}

In this manuscript, we introduced a new method for optimizing model weights 
in a linear pooled forecast ensemble. The method, which is based on 
optimization of the CRPS using a Gibbs posterior distribution, provides 
uncertainty quantification for optimal linear pool weights and allows for a 
prior distribution to inform and influence weight selection. Included with 
the development of the SGP is a theorem for the asymptotic consistency of 
the Gibbs posterior. The SGP was used in three simulation studies and a data 
analysis on the 2023-24 FluSight forecasts to evaluate performance for model 
weighting, and it was compared with two posterior model averaging methods 
and an equally weighted linear pool ensemble. Performance of the methods was 
determined by forecast performance in terms of minimizing the CRPS and LogS, 
and in the second and third simulation studies 
by analyzing the uniformity of the PIT. 
In the simulation studies and the flu forecast analysis, the SGP showed 
superior forecast 
performance when compared to the AVS, BMA, and EQW. In the third simulation 
study, a more informative prior distribution showed the value of regularization
of ensemble weights to be equal.

The promising results show that the SGP is a valuable contribution to the 
field of probabilistic forecast combination. The content of the SGP explored 
in this paper was primarily directed at forecasting a future event, but in 
a scenario where ensemble component model uncertainty is of more concern, 
further exploration 
of the Gibbs posterior should be carried out. This may include further 
exploration into the selection or design of objective prior distributions 
\cite[]{giummole2019objective}. 
Other aspects of the SGP to be explored could be its use on scoring rules 
other than the CRPS or in nonlinear model combination methods. While the 
CRPS is a popular and useful measure of probabilistic forecast performance 
and one which we show is often simple to estimate in the continuous mixture 
distribution case, other scoring rules exist and different rules may be 
preferable for use in different applications. The WIS, for example, remains 
the preferred scoring method of the quantile forecasts for FluSight. An 
adaption of the SGP where the empirical risk is based on the WIS instead of 
the CRPS may be better suited for FluSight and other forecast hubs. Likewise, 
optimizing the linear pool forecast model in (\ref{eq:linp_forc}) may not 
be the ideal method for ensemble building. In fact the linear pool, while 
relatively simple to construct, has its drawbacks such as the difficulty of 
producing a completely uniform PIT \cite[]{gneiting2013combining}. 
Applications of the SGP in the context of the so called generalized linear 
pool would be another direction of further development.

A limitation of this work is the asymptotic consistency theory of the SGP. 
Theorem \ref{thm:sgp_cons} is limited to the case where the data is i.i.d. 
and where the models of the linear pool are fixed. An issue with posterior 
consistency in the dynamic setting is the tendency for the skill of 
probabilistic forecast models to vary in time making a single risk minimizer 
unlikely to exist without strict assumptions. Another limitation, present in 
the flu forecast application, is that forecasts are often missing. In the 
analysis in section \ref{sec:flu_analysis}, only models with 100\% submission 
over the course of the flu season for a given region were included in 
the ensemble forecasts. Over the 
season, many forecast teams failed to submit forecasts during some weeks, 
and even if their forecasts were highly skilled these could not be included 
in the SGP 
as it is developed here. Modifying the SGP to account for missing forecasts 
so that they may factor into an ensemble when included, or by considering 
any of the above recommendations could improve the overall performance 
and/or lead to additional developments.

\section{Acknowledgments}
This work is partially supported by the National Science Foundation under 
Grant No. 2152117. Any opinions, findings, and conclusions or recommendations 
expressed in this material are those of the author(s) and do not necessarily 
reflect the views of the National Science Foundation.

\bibliographystyle{apalike}
\bibliography{master_bib}

\begin{thebibliography}{}

\bibitem[Allen et~al., 2008]{allen2008mathematical}
Allen, L.~J., Brauer, F., Van~den Driessche, P., and Wu, J. (2008).
\newblock {\em {Mathematical Epidemiology}}, volume 1945.
\newblock Springer.

\bibitem[Bassetti et~al., 2018]{bassetti2018bayesian}
Bassetti, F., Casarin, R., and Ravazzolo, F. (2018).
\newblock Bayesian nonparametric calibration and combination of predictive distributions.
\newblock {\em Journal of the American Statistical Association}, 113(522):675--685.

\bibitem[Bernardo and Smith, 1994]{bernardo1994bayesian}
Bernardo, J.~M. and Smith, A.~F. (1994).
\newblock {\em Bayesian Theory}.
\newblock John Wiley \& Sons.

\bibitem[Berrisch and Ziel, 2023]{berrisch2023crps}
Berrisch, J. and Ziel, F. (2023).
\newblock {CRPS learning}.
\newblock {\em Journal of Econometrics}, 237(2):105221.

\bibitem[Biggerstaff et~al., 2016]{biggerstaff2016results}
Biggerstaff, M., Alper, D., Dredze, M., Fox, S., Fung, I. C.-H., Hickmann, K.~S., Lewis, B., Rosenfeld, R., Shaman, J., Tsou, M.-H., et~al. (2016).
\newblock {Results from the Centers for Disease Control and Prevention’s predict the 2013--2014 Influenza Season Challenge}.
\newblock {\em BMC Infectious Diseases}, 16(1):1--10.

\bibitem[Billio et~al., 2013]{billio2013time}
Billio, M., Casarin, R., Ravazzolo, F., and Van~Dijk, H.~K. (2013).
\newblock Time-varying combinations of predictive densities using nonlinear filtering.
\newblock {\em Journal of Econometrics}, 177(2):213--232.

\bibitem[Bissiri et~al., 2016]{bissiri2016general}
Bissiri, P.~G., Holmes, C.~C., and Walker, S.~G. (2016).
\newblock A general framework for updating belief distributions.
\newblock {\em Journal of the Royal Statistical Society Series B: Statistical Methodology}, 78(5):1103--1130.

\bibitem[Boos and Stefanski, 2013]{boos2013essential}
Boos, D.~D. and Stefanski, L.~A. (2013).
\newblock {\em Essential Statistical Inference: Theory and Methods}.
\newblock Springer, New York.

\bibitem[Bracher et~al., 2021]{bracher2021evaluating}
Bracher, J., Ray, E.~L., Gneiting, T., and Reich, N.~G. (2021).
\newblock Evaluating epidemic forecasts in an interval format.
\newblock {\em PLOS Computational Biology}, 17(2):e1010592.

\bibitem[CDC, 2024]{cdc2024previous}
CDC (2024).
\newblock {Centers for Disease Control and Prevention FluSight: Flu Forecasting}.
\newblock \url{https://www.cdc.gov/flu-forecasting/data-vis/index.html}.
\newblock Accessed: 2024-09-24.

\bibitem[Claeskens et~al., 2016]{claeskens2016forecast}
Claeskens, G., Magnus, J.~R., Vasnev, A.~L., and Wang, W. (2016).
\newblock The forecast combination puzzle: A simple theoretical explanation.
\newblock {\em International Journal of Forecasting}, 32(3):754--762.

\bibitem[Clemen, 1989]{clemen1989combining}
Clemen, R.~T. (1989).
\newblock Combining forecasts: A review and annotated bibliography.
\newblock {\em International Journal of Forecasting}, 5(4):559--583.

\bibitem[Clyde and Iversen, 2013]{clyde2013bayesian}
Clyde, M. and Iversen, E.~S. (2013).
\newblock {Bayesian model averaging in the M-open framework}.
\newblock In Damien, P., Dellaportas, P., Polson, N.~G., and Stephens, D.~A., editors, {\em Bayesian Theory and Applications}, page 483–498. Oxford University Press.

\bibitem[Collins, 2007]{collins2007ensembles}
Collins, M. (2007).
\newblock Ensembles and probabilities: a new era in the prediction of climate change.
\newblock {\em Philosophical Transactions of the Royal Society A: Mathematical, Physical and Engineering Sciences}, 365(1857):1957--1970.

\bibitem[Cramer et~al., 2022]{Cramer2022-hub-dataset}
Cramer, E.~Y., Huang, Y., Wang, Y., Ray, E.~L., Cornell, M., Bracher, J., Brennen, A., Castro~Rivadeneira, A.~J., Gerding, A., House, K., Jayawardena, D., Kanji, A.~H., Khandelwal, A., Le, K., Niemi, J., Stark, A., Shah, A., Wattanachit, N., Zorn, M.~W., Reich, N.~G., and Consortium, U. C.-. F.~H. (2022).
\newblock {The United States COVID-19 forecast hub dataset}.
\newblock {\em Scientific Data}, 9(1):462.

\bibitem[Farrow et~al., 2017]{farrow2017human}
Farrow, D.~C., Brooks, L.~C., Hyun, S., Tibshirani, R.~J., Burke, D.~S., and Rosenfeld, R. (2017).
\newblock A human judgment approach to epidemiological forecasting.
\newblock {\em PLOS Computational Biology}, 13(3):e1005248.

\bibitem[Fox et~al., 2024]{fox2024optimizing}
Fox, S.~J., Kim, M., Meyers, L.~A., Reich, N.~G., and Ray, E.~L. (2024).
\newblock Optimizing disease outbreak forecast ensembles.
\newblock {\em Emerging Infectious Diseases}, 30(9):1967.

\bibitem[Frazier et~al., 2023]{frazier2023solving}
Frazier, D.~T., Covey, R., Martin, G.~M., and Poskitt, D. (2023).
\newblock Solving the forecast combination puzzle.
\newblock {\em arXiv preprint arXiv:2308.05263}.

\bibitem[Gabry et~al., 2022]{gabry2022stan}
Gabry, J., Češnovar, R., and Johnson, A. (2022).
\newblock {\em cmdstanr: R Interface to 'CmdStan'}.
\newblock https://mc-stan.org/cmdstanr/, https://discourse.mc-stan.org.

\bibitem[Gebetsberger et~al., 2018]{gebetsberger2018estimation}
Gebetsberger, M., Messner, J.~W., Mayr, G.~J., and Zeileis, A. (2018).
\newblock Estimation methods for nonhomogeneous regression models: Minimum continuous ranked probability score versus maximum likelihood.
\newblock {\em Monthly Weather Review}, 146(12):4323--4338.

\bibitem[Gelman et~al., 2013]{gelman2013bayesian}
Gelman, A., Carlin, J.~B., Stern, H.~S., Dunson, D.~B., Vehtari, A., and Rubin, D.~B. (2013).
\newblock {\em Bayesian Data Analysis, Third Edition}.
\newblock Chapman and Hall/CRC.

\bibitem[Geweke and Amisano, 2011]{geweke2011optimal}
Geweke, J. and Amisano, G. (2011).
\newblock Optimal prediction pools.
\newblock {\em Journal of Econometrics}, 164(1):130--141.

\bibitem[Github, 2024]{mathis2023flusight}
Github (2024).
\newblock {FluSight-forecast-hub}.
\newblock \url{https://github.com/cdcepi/FluSight-forecast-hub}.
\newblock Accessed: 2024-10-22.

\bibitem[Giummol{\`e} et~al., 2019]{giummole2019objective}
Giummol{\`e}, F., Mameli, V., Ruli, E., and Ventura, L. (2019).
\newblock Objective bayesian inference with proper scoring rules.
\newblock {\em Test}, 28(3):728--755.

\bibitem[Gneiting et~al., 2007]{gneiting2007probabilistic}
Gneiting, T., Balabdaoui, F., and Raftery, A.~E. (2007).
\newblock Probabilistic forecasts, calibration and sharpness.
\newblock {\em Journal of the Royal Statistical Society Series B: Statistical Methodology}, 69(2):243--268.

\bibitem[Gneiting and Katzfuss, 2014]{gneiting2014probabilistic}
Gneiting, T. and Katzfuss, M. (2014).
\newblock Probabilistic forecasting.
\newblock {\em Annual Review of Statistics and Its Application}, 1:125--151.

\bibitem[Gneiting and Raftery, 2007]{gneiting2007strictly}
Gneiting, T. and Raftery, A.~E. (2007).
\newblock Strictly proper scoring rules, prediction, and estimation.
\newblock {\em Journal of the American Statistical Association}, 102(477):359--378.

\bibitem[Gneiting et~al., 2005]{gneiting2005calibrated}
Gneiting, T., Raftery, A.~E., Westveld, A.~H., and Goldman, T. (2005).
\newblock {Calibrated probabilistic forecasting using ensemble model output statistics and minimum CRPS estimation}.
\newblock {\em Monthly Weather Review}, 133(5):1098--1118.

\bibitem[Gneiting and Ranjan, 2011]{gneiting2011comparing}
Gneiting, T. and Ranjan, R. (2011).
\newblock Comparing density forecasts using threshold-and quantile-weighted scoring rules.
\newblock {\em Journal of Business \& Economic Statistics}, 29(3):411--422.

\bibitem[Gneiting and Ranjan, 2013]{gneiting2013combining}
Gneiting, T. and Ranjan, R. (2013).
\newblock Combining predictive distributions.
\newblock {\em Electronic Journal of Statistics}, 7:1747.

\bibitem[Gyamerah et~al., 2020]{gyamerah2020probabilistic}
Gyamerah, S.~A., Ngare, P., and Ikpe, D. (2020).
\newblock {Probabilistic forecasting of crop yields via quantile random forest and Epanechnikov kernel function}.
\newblock {\em Agricultural and Forest Meteorology}, 280:107808.

\bibitem[Hall and Mitchell, 2007]{hall2007combining}
Hall, S.~G. and Mitchell, J. (2007).
\newblock Combining density forecasts.
\newblock {\em International Journal of Forecasting}, 23(1):1--13.

\bibitem[Hamill, 2001]{hamill2001interpretation}
Hamill, T.~M. (2001).
\newblock Interpretation of rank histograms for verifying ensemble forecasts.
\newblock {\em Monthly Weather Review}, 129(3):550--560.

\bibitem[Hong et~al., 2016]{hong2016probabilistic}
Hong, T., Pinson, P., Fan, S., Zareipour, H., Troccoli, A., and Hyndman, R.~J. (2016).
\newblock Probabilistic energy forecasting: Global energy forecasting competition 2014 and beyond.
\newblock {\em International Journal of Forecasting}, 32(3):896--913.

\bibitem[Hong et~al., 2020]{hong2020energy}
Hong, T., Pinson, P., Wang, Y., Weron, R., Yang, D., and Zareipour, H. (2020).
\newblock Energy forecasting: A review and outlook.
\newblock {\em IEEE Open Access Journal of Power and Energy}, 7:376--388.

\bibitem[Hyndman, 2020]{hyndman2020brief}
Hyndman, R.~J. (2020).
\newblock A brief history of forecasting competitions.
\newblock {\em International Journal of Forecasting}, 36(1):7--14.

\bibitem[Hyndman and Khandakar, 2008]{hyndman2008automatic}
Hyndman, R.~J. and Khandakar, Y. (2008).
\newblock Automatic time series forecasting: the forecast package for {R}.
\newblock {\em Journal of Statistical Software}, 27(3):1--22.

\bibitem[Jiang and Tanner, 2008]{jiang2008gibbs}
Jiang, W. and Tanner, M.~A. (2008).
\newblock Gibbs posterior for variable selection in high-dimensional classification and data mining.
\newblock {\em The Annals of Statistics}, 36(5):2207--2231.

\bibitem[Joslyn and LeClerc, 2012]{joslyn2012uncertainty}
Joslyn, S.~L. and LeClerc, J.~E. (2012).
\newblock Uncertainty forecasts improve weather-related decisions and attenuate the effects of forecast error.
\newblock {\em Journal of Experimental Psychology: Applied}, 18(1):126.

\bibitem[Kapetanios et~al., 2015]{kapetanios2015generalised}
Kapetanios, G., Mitchell, J., Price, S., and Fawcett, N. (2015).
\newblock Generalised density forecast combinations.
\newblock {\em Journal of Econometrics}, 188(1):150--165.

\bibitem[Kermack and McKendrick, 1927]{kermack1927contribution}
Kermack, W.~O. and McKendrick, A.~G. (1927).
\newblock A contribution to the mathematical theory of epidemics.
\newblock {\em Proceedings of the Royal Society of London. Series A, Containing papers of a mathematical and physical character}, 115(772):700--721.

\bibitem[Koop and Korobilis, 2013]{koop2013large}
Koop, G. and Korobilis, D. (2013).
\newblock {Large time-varying parameter VARs}.
\newblock {\em Journal of Econometrics}, 177(2):185--198.

\bibitem[Lavine et~al., 2021]{lavine2021adaptive}
Lavine, I., Lindon, M., and West, M. (2021).
\newblock Adaptive variable selection for sequential prediction in multivariate dynamic models.
\newblock {\em Bayesian Analysis}, 16(4):1059--1083.

\bibitem[Li et~al., 2023]{li2023bayesian}
Li, L., Kang, Y., and Li, F. (2023).
\newblock Bayesian forecast combination using time-varying features.
\newblock {\em International Journal of Forecasting}, 39(3):1287--1302.

\bibitem[Li et~al., 2019]{li2019combining}
Li, T., Wang, Y., and Zhang, N. (2019).
\newblock Combining probability density forecasts for power electrical loads.
\newblock {\em IEEE Transactions on Smart Grid}, 11(2):1679--1690.

\bibitem[Lichtendahl~Jr et~al., 2013]{lichtendahl2013better}
Lichtendahl~Jr, K.~C., Grushka-Cockayne, Y., and Winkler, R.~L. (2013).
\newblock Is it better to average probabilities or quantiles?
\newblock {\em Management Science}, 59(7):1594--1611.

\bibitem[Loaiza-Maya et~al., 2021]{loaiza2021focused}
Loaiza-Maya, R., Martin, G.~M., and Frazier, D.~T. (2021).
\newblock {Focused Bayesian prediction}.
\newblock {\em Journal of Applied Econometrics}, 36(5):517--543.

\bibitem[Makridakis et~al., 2020]{makridakis2020m4}
Makridakis, S., Spiliotis, E., and Assimakopoulos, V. (2020).
\newblock The {M4} competition: 100,000 time series and 61 forecasting methods.
\newblock {\em International Journal of Forecasting}, 36(1):54--74.

\bibitem[Martin and Syring, 2022]{martin2022direct}
Martin, R. and Syring, N. (2022).
\newblock Direct gibbs posterior inference on risk minimizers: Construction, concentration, and calibration.
\newblock In {\em Handbook of Statistics}, volume~47, pages 1--41. Elsevier.

\bibitem[Mathis et~al., 2024]{mathis2024evaluation}
Mathis, S.~M., Webber, A.~E., Le{\'o}n, T.~M., Murray, E.~L., Sun, M., White, L.~A., Brooks, L.~C., Green, A., Hu, A.~J., Rosenfeld, R., et~al. (2024).
\newblock Evaluation of {FluSight} influenza forecasting in the 2021--22 and 2022--23 seasons with a new target laboratory-confirmed influenza hospitalizations.
\newblock {\em Nature Communications}, 15(1):6289.

\bibitem[McAlinn and West, 2019]{mcalinn2019dynamic}
McAlinn, K. and West, M. (2019).
\newblock {Dynamic Bayesian predictive synthesis in time series forecasting}.
\newblock {\em Journal of Econometrics}, 210(1):155--169.

\bibitem[McAndrew and Reich, 2021]{mcandrew2021adaptively}
McAndrew, T. and Reich, N.~G. (2021).
\newblock Adaptively stacking ensembles for influenza forecasting.
\newblock {\em Statistics in Medicine}, 40(30):6931--6952.

\bibitem[Morgan et~al., 2018]{morgan2018wisdom}
Morgan, J.~J., Wilson, O.~C., and Menon, P.~G. (2018).
\newblock The wisdom of crowds approach to influenza-rate forecasting.
\newblock In {\em ASME international mechanical engineering congress and exposition}, volume 52026, page V003T04A048. American Society of Mechanical Engineers.

\bibitem[Newey and McFadden, 1994]{newey1994large}
Newey, W.~K. and McFadden, D. (1994).
\newblock Large sample estimation and hypothesis testing.
\newblock {\em Handbook of Econometrics}, 4:2111--2245.

\bibitem[Osthus et~al., 2019]{osthus2019dynamic}
Osthus, D., Gattiker, J., Priedhorsky, R., and Del~Valle, S.~Y. (2019).
\newblock {Dynamic Bayesian influenza forecasting in the United States with hierarchical discrepancy (with discussion)}.
\newblock {\em Bayesian Analysis}, 14(1):261--312.

\bibitem[Osthus and Moran, 2021]{osthus2021multiscale}
Osthus, D. and Moran, K.~R. (2021).
\newblock Multiscale influenza forecasting.
\newblock {\em Nature Communications}, 12(1):2991.

\bibitem[Palmer, 2002]{palmer2002economic}
Palmer, T.~N. (2002).
\newblock The economic value of ensemble forecasts as a tool for risk assessment: From days to decades.
\newblock {\em Quarterly Journal of the Royal Meteorological Society: A journal of the atmospheric sciences, applied meteorology and physical oceanography}, 128(581):747--774.

\bibitem[Raftery, 1996]{raftery1996hypothesis}
Raftery, A.~E. (1996).
\newblock Hypothesis testing and model selection.
\newblock In Gilks, W., Richardson, S., and Spiegelhalter, D., editors, {\em Markov Chain Monte Carlo in Practice: Interdisciplinary Statistics}, pages 163--187. Chapman and Hall.

\bibitem[Raftery et~al., 2010]{raftery2010online}
Raftery, A.~E., K{\'a}rn{\`y}, M., and Ettler, P. (2010).
\newblock Online prediction under model uncertainty via dynamic model averaging: Application to a cold rolling mill.
\newblock {\em Technometrics}, 52(1):52--66.

\bibitem[Ramos et~al., 2013]{ramos2013probabilistic}
Ramos, M.~H., Van~Andel, S.~J., and Pappenberger, F. (2013).
\newblock Do probabilistic forecasts lead to better decisions?
\newblock {\em Hydrology and Earth System Sciences}, 17(6):2219--2232.

\bibitem[Ranjan and Gneiting, 2010]{ranjan2010combining}
Ranjan, R. and Gneiting, T. (2010).
\newblock Combining probability forecasts.
\newblock {\em Journal of the Royal Statistical Society Series B: Statistical Methodology}, 72(1):71--91.

\bibitem[Reich et~al., 2019a]{reich2019collaborative}
Reich, N.~G., Brooks, L.~C., Fox, S.~J., Kandula, S., McGowan, C.~J., Moore, E., Osthus, D., Ray, E.~L., Tushar, A., Yamana, T.~K., et~al. (2019a).
\newblock A collaborative multiyear, multimodel assessment of seasonal influenza forecasting in the united states.
\newblock {\em Proceedings of the National Academy of Sciences}, 116(8):3146--3154.

\bibitem[Reich et~al., 2019b]{reich2019collaborativeens}
Reich, N.~G., McGowan, C.~J., Yamana, T.~K., Tushar, A., Ray, E.~L., Osthus, D., Kandula, S., Brooks, L.~C., Crawford-Crudell, W., Gibson, G.~C., et~al. (2019b).
\newblock A collaborative multi-model ensemble for real-time influenza season forecasting in the us.
\newblock {\em bioRxiv}, page 566604.

\bibitem[Smith and Wallis, 2009]{smith2009simple}
Smith, J. and Wallis, K.~F. (2009).
\newblock A simple explanation of the forecast combination puzzle.
\newblock {\em Oxford Bulletin of Economics and Statistics}, 71(3):331--355.

\bibitem[{Stan Development Team}, 2024]{stan2024manual}
{Stan Development Team} (2024).
\newblock Stan modeling language users guide and reference manual, 2.34.
\newblock \url{https://mc-stan.org}.
\newblock Accessed: 2024-10-22.

\bibitem[Stone, 1961]{stone1961opinion}
Stone, M. (1961).
\newblock The opinion pool.
\newblock {\em The Annals of Mathematical Statistics}, 32:1339--1342.

\bibitem[Syring and Martin, 2017]{syring2017gibbs}
Syring, N. and Martin, R. (2017).
\newblock Gibbs posterior inference on the minimum clinically important difference.
\newblock {\em Journal of Statistical Planning and Inference}, 187:67--77.

\bibitem[Tallman and West, 2024]{tallman2024bayesian}
Tallman, E. and West, M. (2024).
\newblock Bayesian predictive decision synthesis.
\newblock {\em Journal of the Royal Statistical Society Series B: Statistical Methodology}, 86(2):340--363.

\bibitem[Thorey et~al., 2017]{thorey2017online}
Thorey, J., Mallet, V., and Baudin, P. (2017).
\newblock Online learning with the continuous ranked probability score for ensemble forecasting.
\newblock {\em Quarterly Journal of the Royal Meteorological Society}, 143(702):521--529.

\bibitem[Ulloa, 2019]{ulloa2019}
Ulloa, N. (2019).
\newblock {\em Bayesian hierarchical modeling for disease outbreaks}.
\newblock PhD thesis, Iowa State University Department of Statistics.

\bibitem[Van~der Vaart, 1998]{van1998asymptotic}
Van~der Vaart, A.~W. (1998).
\newblock {\em Asymptotic Statistics}.
\newblock Cambridge University Press, Cambridge, United Kingdom.

\bibitem[Vehtari et~al., 2021]{vehtari2021rank}
Vehtari, A., Gelman, A., Simpson, D., Carpenter, B., and B{\"u}rkner, P.-C. (2021).
\newblock {Rank-normalization, folding, and localization: An improved {\^R} for assessing convergence of MCMC (with discussion)}.
\newblock {\em Bayesian Analysis}, 16(2):667--718.

\bibitem[Wadsworth and Niemi, 2025a]{wadsworth2025forecasting}
Wadsworth, S. and Niemi, J. (2025a).
\newblock Forecasting influenza hospitalizations using a bayesian hierarchical nonlinear model with discrepancy.
\newblock {\em arXiv preprint arXiv:2412.14339}.

\bibitem[Wadsworth and Niemi, 2025b]{wadsworth2025quantile}
Wadsworth, S. and Niemi, J. (2025b).
\newblock {Quantile forecast matching with a Bayesian quantile Gaussian process model}.
\newblock {\em arXiv preprint arXiv:2502.06605}.

\bibitem[Wadsworth et~al., 2023]{wadsworth2023mixture}
Wadsworth, S., Niemi, J., and Reich, N. (2023).
\newblock Mixture distributions for probabilistic forecasts of disease outbreaks.
\newblock {\em arXiv preprint arXiv:2310.11939}.

\bibitem[Wang et~al., 2023]{wang2023forecast}
Wang, X., Hyndman, R.~J., Li, F., and Kang, Y. (2023).
\newblock Forecast combinations: An over 50-year review.
\newblock {\em International Journal of Forecasting}, 39(4):1518--1547.

\bibitem[Widgren et~al., 2019]{widgren2019siminf}
Widgren, S., Bauer, P., Eriksson, R., and Engblom, S. (2019).
\newblock {SimInf}: An {R} package for data-driven stochastic disease spread simulations.
\newblock {\em Journal of Statistical Software}, 91(12):1--42.

\bibitem[Yao et~al., 2018]{yao2018using}
Yao, Y., Vehtari, A., Simpson, D., and Gelman, A. (2018).
\newblock {Using stacking to average Bayesian predictive distributions (with discussion)}.
\newblock {\em Bayesian Analysis}, 13(3):917--1003.

\bibitem[Zamo and Naveau, 2018]{zamo2018estimation}
Zamo, M. and Naveau, P. (2018).
\newblock Estimation of the continuous ranked probability score with limited information and applications to ensemble weather forecasts.
\newblock {\em Mathematical Geosciences}, 50(2):209--234.

\bibitem[Zhang, 2006]{zhang2006information}
Zhang, T. (2006).
\newblock Information-theoretic upper and lower bounds for statistical estimation.
\newblock {\em IEEE Transactions on Information Theory}, 52(4):1307--1321.

\end{thebibliography}

%%%%
% Reference section comes before the appendix

%\bibliographystyle{acm} % use for numbered citations along with options given in preamble. Look at the main thesis.tex file
% \bibliographystyle{apa}
% \bibliography{master_bib}

% \jarad{Check all references for accuracy, capitalization.}
% %\section{Bibliography}
% \bibliographystyle{apa}
% % \vspace{-20pt}
% \begingroup
%     \setlength{\bibsep}{13.2pt}
%     \linespread{1}\selectfont
%     \bibliography{master_bib}
% \endgroup
% \clearpage
% \pagebreak

%%%%%
%% Appendix
% This section may or may not be included
% Chapter 2 shows a double appendix example. Please use A or B as in Appendix A if there are multiple appendix. Use "Appendix A:" before writing the title
% Chapter 3 shows a single appendix example. Use "Appendix:" before writing the title 
\section{Appendix: Proofs of equation (\ref{eq:crps_ens_mse}) and theorem 
\ref{thm:sgp_cons}}

\subsection{Proof of equation (\ref{eq:crps_ens_mse})} \label{app:crps_ens_pf}

\begin{proof}
The random variables $X$ and $X'$ in (\ref{eq:crps_mse}) are independent 
copies of the same random variable with mixture distribution PDF 
\[
\bar{p}(x) = \sum_{c = 1}^C w_c p_c(x)
\]
where each $p_c(x)$ is a PDF of a continuous random variable. The random 
variables $X_c$ and $X'_c$ are independent copies of a random variable with 
PDF $p_c(x)$. Then the second expectation in (\ref{eq:crps_mse}) is  
    \begin{flalign*}
        E|X - X'| &= \int_{\mathcal{X}} \int_{\mathcal{X}} |x - x'| p(x)p(x') 
        dxdx' \\
        &= \int_{\mathcal{X}} \int_{\mathcal{X}} |x - x'| 
        \sum_{i = 1}^C w_c p_c(x) \sum_{d = 1}^C w_d p_d(x') dx dx' \\
        &= \sum_{c = 1}^C \sum_{d = 1}^C w_cw_d \int_{\mathcal{X}} 
        \int_{\mathcal{X}} |x-x'|p_c(x)p_d(x') dx dx' \\
        &= \sum_{c = 1}^C \sum_{d = 1}^C w_cw_d E|X_c-X'_d|
    \end{flalign*}
Similary for the first expectation
\[
    E|X - y| = \sum_{c = 1}^C w_c E|X_c - y|
\]
and the result in (\ref{eq:crps_ens_mse}) follows.
\end{proof}

\subsection{Proof of theorem \ref{thm:sgp_cons}} \label{app:sgp_cons_pf}
\begin{proof}
\cite{martin2022direct} identify three sufficient conditions needed to prove 
Gibbs posterior consistency. The first is that the prior distribution give 
sufficient mass to the risk minimizer $\omega^*$. This is met with the selected 
Dirichlet prior. The other two conditions are a uniform law of large numbers 
and a separation or identifiability condition. Respectively these two 
conditions are (\ref{eq:cons_ull}) and (\ref{eq:cons_sep}).
\begin{equation}
    \underset{\omega \in \Omega}{\text{sup}} |G_n(\omega) - G(\omega)| 
    \rightarrow 0 \text{ in } \mathcal{L}-\text{probability}
    \label{eq:cons_ull}
\end{equation}

\begin{equation}
    \underset{\omega:d(\omega, \omega^*) > \delta}{\text{inf}} \{G(\omega) - 
    G(\omega^*)\} > 0 \text{ for any } \delta > 0
    \label{eq:cons_sep}
\end{equation}

\noindent We first show (\ref{eq:cons_ull}). By lemma 2.4 in 
\cite{newey1994large}, if 
\begin{enumerate}%[\hspace{.5cm}i)]
    \item the data are iid
    \item $\Omega$ is compact
    \item $CRPS(\bar{P}_{\omega}, y)$ is continuous at each $\omega \in 
    \Omega$ with probability 1
    \item there is a function $m(y)$ such that $|CRPS(\bar{P}_{\omega}, y)| 
    \leq m(y)$ and $E[m(Y)] < \infty$
\end{enumerate}
then $G(\omega)$ is continuous and (\ref{eq:cons_ull}) follows. Conditions i) 
and ii) are met by assumption and iii) is met by the continuity of the CRPS.
To show iv) we let
\begin{align*}
    \text{CRPS}(\bar{P}_{\omega}, y) &= E|X - y| - \frac{1}{2}E|X - X'| \\
    &\leq E|X - y| \\
    &\leq E\left[|X| + |y|\right] \\
    &=E|X| + |y| := m(y)
\end{align*}
Then by the assumption that $E|Y| < \infty$, $E[m(Y)] < \infty$ meeting 
condition iv) and showing (\ref{eq:cons_ull}).

Now to show (\ref{eq:cons_sep}). Consider the open set $B_{\delta'} = 
\{\omega : d(\omega, \omega^*) < \delta'\}$ and the closed set 
$\overline{B}_{\delta} = \{\omega : d(\omega, \omega^*) \leq \delta\}$ 
where $0 < \delta' < \delta$. Since $B_{\delta'}$ is open, the set 
$\Omega \setminus B_{\delta'} \subset \Omega$ is a closed subset of a 
compact set and is thus compact. Because  $G(\cdot)$ is continuous, the 
set $G(\Omega \setminus B_{\delta'})$ is also compact and thus contains 
its infimum. Now $\Omega \setminus \overline{B}_{\delta} \subset \Omega 
\setminus B_{\delta'}$, so $G(\Omega \setminus \overline{B}_{\delta'}) 
\subset G(\Omega \setminus B_{\delta'})$. Thus 
\[
     G(\omega^*) < \underset{\Omega \setminus B_{\delta'}}{\text{inf}} 
     G(\omega) \leq \underset{\Omega \setminus 
     \overline{B}_{\delta}}{\text{inf}} G(\omega)
\]
and (\ref{eq:cons_sep}) holds.
\end{proof}

\section*{Supplemental Material for Bayesian stacking via proper scoring rule
optimization using a Gibbs
posterior}
\begin{figure}[hbt!]
\centering
%\begin{subfigure}{}
  % \centering
  \includegraphics[width=.48\linewidth]{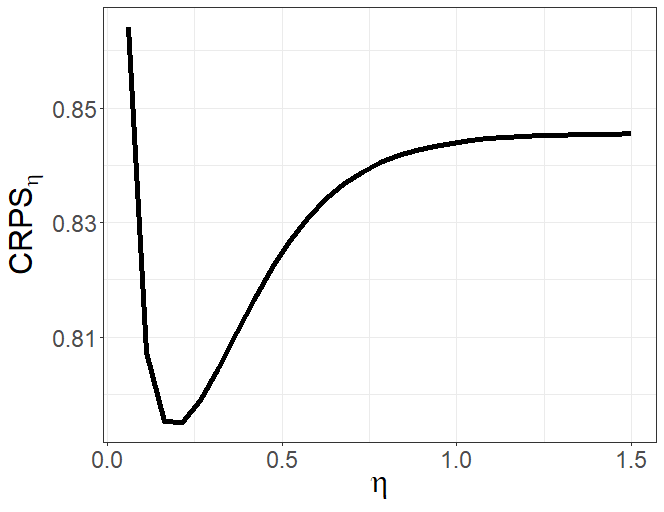}
  % \caption{A subfigure}
  % \label{fig:sub1}
%\end{subfigure}%
%\begin{subfigure}{}
  \centering
  \includegraphics[width=.48\linewidth]{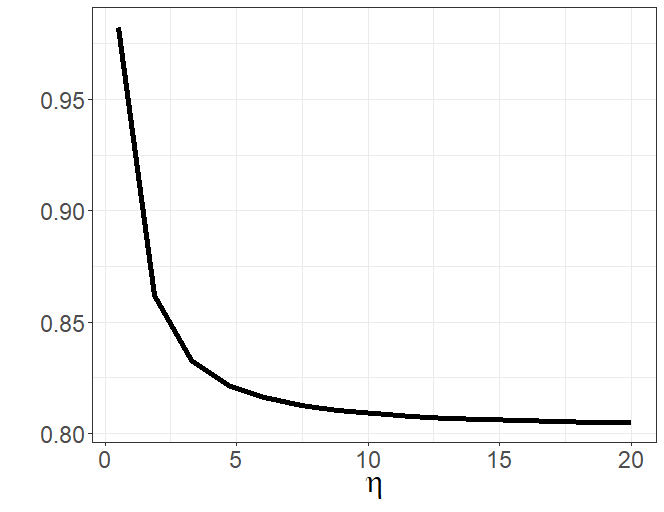}
  % \caption{A subfigure}
  % \label{fig:sub2}
%\end{subfigure}
\caption{An example of CRPS values for different values of $\eta$ after 
optimizing competing model weights for AVS (left) and for SGP (right).}
\label{fig:min_etas}
\end{figure}

\begin{figure}[hbt!]
\centering
%\begin{subfigure}{}
  % \centering
  \includegraphics[width=.48\linewidth]{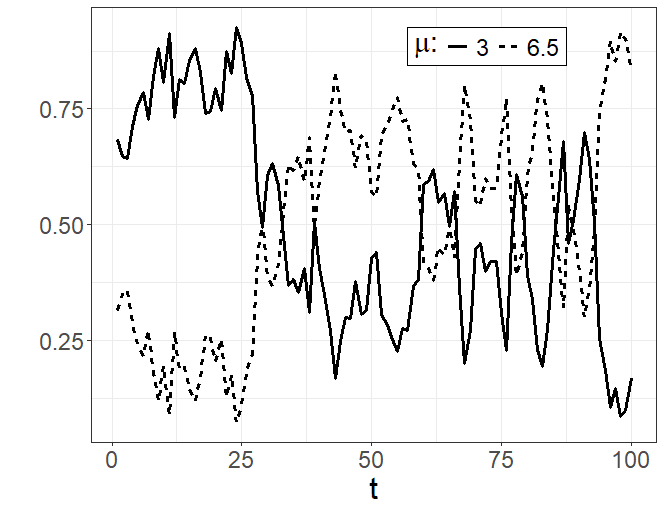}
  % \caption{A subfigure}
  % \label{fig:sub1}
%\end{subfigure}%
\hspace{0.01\textwidth}
%\begin{subfigure}{}
  \centering
  \includegraphics[width=.48\linewidth]{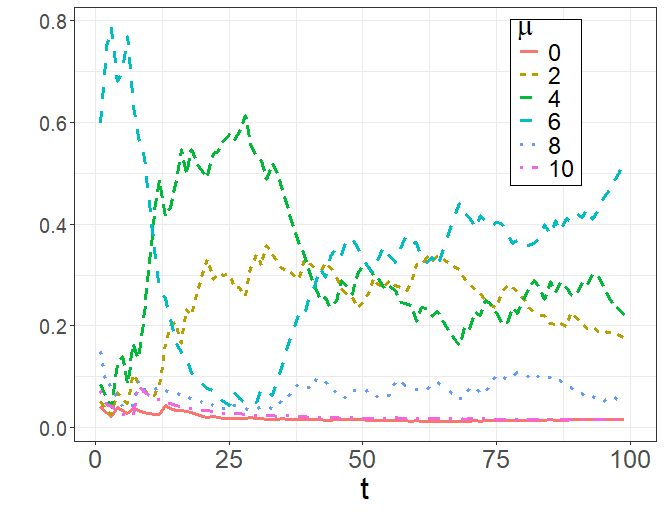}
  % \caption{A subfigure}
  % \label{fig:sub2}
%\end{subfigure}
\caption{An example of simulated component weights for 100 time steps. 
The figure shows the simulated weights for the two components of the true 
distribution (left) and the weights optimized under SGP for the six competing 
component models (right).}
\label{fig:dyn_wts_ex}
\end{figure}

\begin{figure}[hbt!]
\centering
  \includegraphics[width=.98\linewidth]{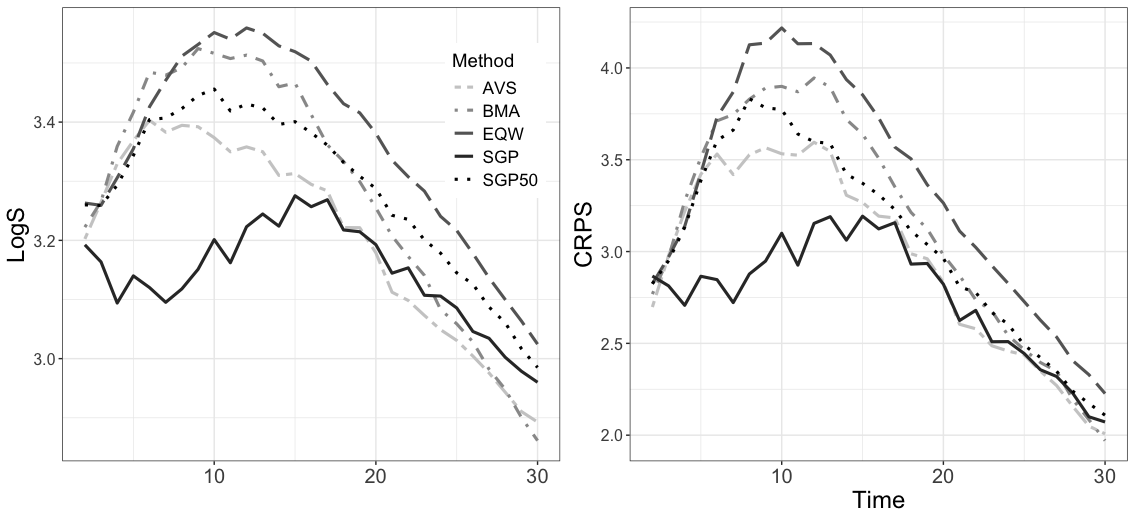}
\caption{Plots showing results for one step ahead forecasts for SIR
data for the four
weighting methods SGP, AVS, BMA, and EQW. The median LogS (left) and 
CRPS (right) over 3,000 replicates of one step
ahead predictions at 29 time points colored by weighting methods.
}
\label{fig:sir_sim_med}
\end{figure}

\begin{figure}[hbt!]
\centering

  \includegraphics[width=.6\linewidth]{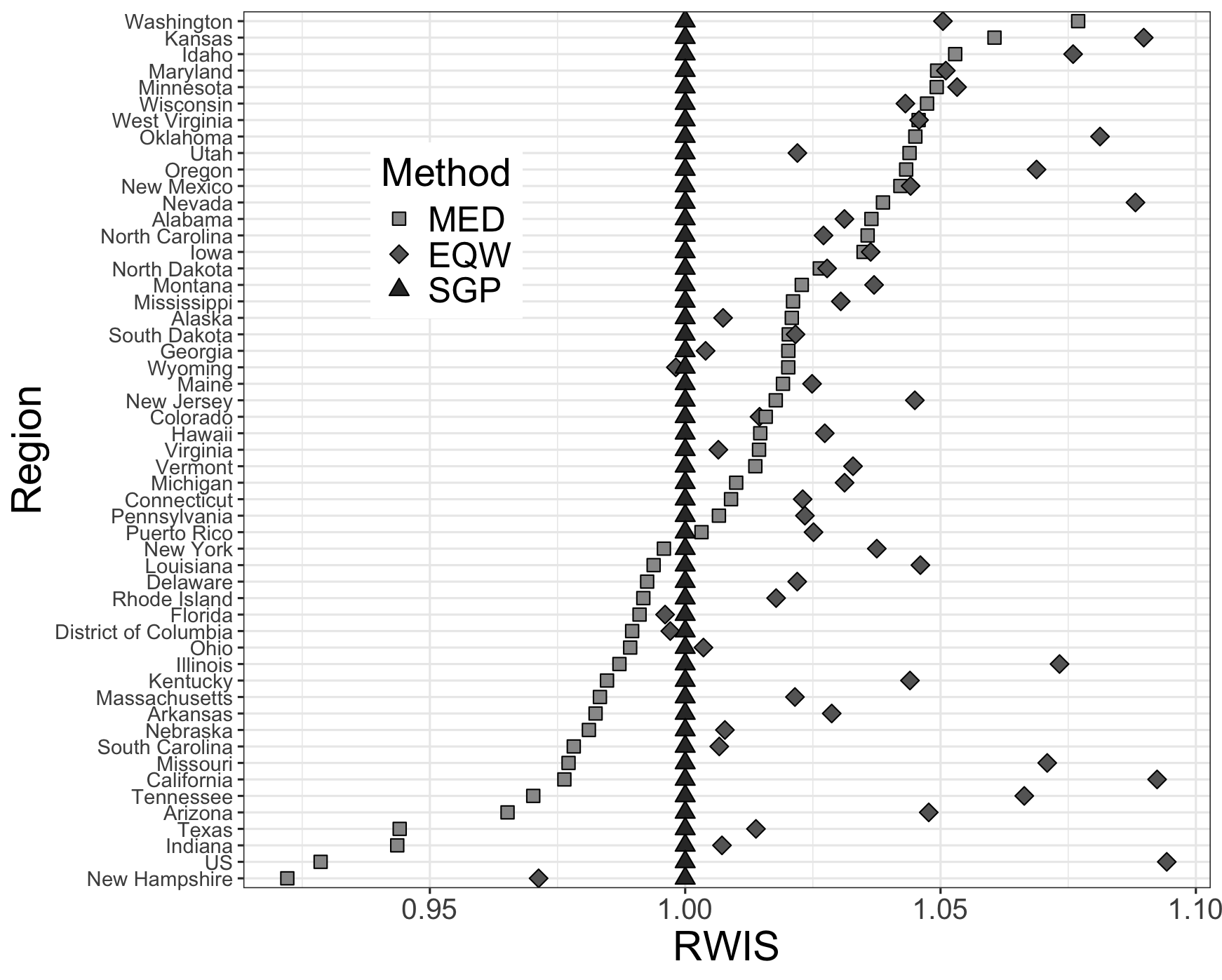}
\caption{
Mean over regions WIS for the four ensemble methods divided
by mean WIS for the SGP
separated by shape and color.}
\label{fig:flu_wis}
\end{figure}

\begin{table}[h!]
\centering
\caption{Table showing the number of times ensemble forecasts for each method, 
SGP, AVS, BMA, and EQW, were ranked 1st, 2nd, 3rd, and 4th according to mean 
CRPS by region or week. There were 53
regions evaluated and 29 weeks, so each column under region should sum to 53
and each column under week to 29.}
 \begin{tabular}{cccccccc}
 % {@{\extracolsep{\fill}}
 %    l
 %    *{8}{c}
 %    S[table-format=4.0]}
 \\
  Ranking
  & \multicolumn{3}{c}{By region CRPS}
  &
  & \multicolumn{3}{c}{By week CRPS} \\
  \cmidrule{2-4}  \cmidrule{6-8}
  & MED & EQW & SGP & & MED & EQW & SGP\\
  \midrule
  $1^{st}$ & 21 & 1  & \textbf{31} & & 11 & 1  & \textbf{17}\\
  $2^{nd}$ & 22 & 12 & 19          & & 15 & 9  & 5\\
  $3^{rd}$ & 10 & 40 & 3           & & 3  & 19 & 7\\
  \bottomrule
 \end{tabular}
 \label{tab:forecast_rank_wis}
\end{table}

% \bibliographystyle{plainnat}
% \bibliography{master_bib}  %%% Uncomment this line and comment out the ``thebibliography'' section below to use the external .bib file (using bibtex) .

%%% Uncomment this section and comment out the \bibliography{references} line above to use inline references.
% \begin{thebibliography}{1}

% 	\bibitem{kour2014real}
% 	George Kour and Raid Saabne.
% 	\newblock Real-time segmentation of on-line handwritten arabic script.
% 	\newblock In {\em Frontiers in Handwriting Recognition (ICFHR), 2014 14th
% 			International Conference on}, pages 417--422. IEEE, 2014.

% 	\bibitem{kour2014fast}
% 	George Kour and Raid Saabne.
% 	\newblock Fast classification of handwritten on-line arabic characters.
% 	\newblock In {\em Soft Computing and Pattern Recognition (SoCPaR), 2014 6th
% 			International Conference of}, pages 312--318. IEEE, 2014.

% 	\bibitem{hadash2018estimate}
% 	Guy Hadash, Einat Kermany, Boaz Carmeli, Ofer Lavi, George Kour, and Alon
% 	Jacovi.
% 	\newblock Estimate and replace: A novel approach to integrating deep neural
% 	networks with existing applications.
% 	\newblock {\em arXiv preprint arXiv:1804.09028}, 2018.

% \end{thebibliography}

% \input{sup_man}
\end{document}